\documentclass[pdflatex,sn-basic,iicol]{sn-jnl}% Basic Springer Nature Reference Style/Chemistry Reference Style
\usepackage{url}
\usepackage{float}
\usepackage{supertabular}
\usepackage{graphicx}
\DeclareRobustCommand{\ion}[2]{%
\relax\ifmmode
\ifx\testbx\f@series
{\mathbf{#1\,\mathsc{#2}}}\else
{\mathrm{#1\,\mathsc{#2}}}\fi
\else\textup{#1\,{\mdseries\textsc{#2}}}%
\fi}
\newcommand{\bz}{$\langle B_z \rangle$}

\newcommand{\kms}{km\,s$^{-1}$}

\newcommand{\apjl}{$ApJL$}

\newcommand{\aap}{A\&A}

\newcommand{\apjs}{ApJS}

\RequirePackage{color}

\definecolor{adu}{rgb}{0.0, 0.1, 0.7}

\begin{document}
%\nolinenumbers

%\title[UV Spectropolarimetry of Hot Star Magnetospheres]{Ultraviolet Spectropolarimetry: Hot Star Magnetospheres}

\title{Ultraviolet Spectropolarimetric Diagnostics of Hot Star Magnetospheres}

\author*[]{\fnm{A. }\sur{ud-Doula$^{1}$}}\email{asif@psu.edu}
%\affiliation{Penn State Scranton, 120 Ridge View Drive, Dunmore, PA 18512, US}
%\affiliation{Department of Physics and Astronomy, University of Delaware, 217 Sharp Lab, Newark, Delaware, 19716, USA}
%\and
%\author{R. Casini$^{2}$}
%\affiliation{High Altitude Observatory, National Center for Atmospheric Research, P.O. Box 3000, Boulder CO 80307-3000, USA}

\author{M. C. M. Cheung$^{2}$}
%\affiliation{Lockheed Martin Solar and Astrophysics Laboratory, 3251 Hanover St, Palo Alto, CA 94304, USA}

\author{A. David-Uraz$^{3,4}$}
%\affiliation{Department of Physics and Astronomy, Howard University, Washington, DC 20059, USA}
%\affiliation{Center for Research and Exploration in Space Science and Technology, and X-ray Astrophysics Laboratory, NASA/GSFC, Greenbelt, MD 20771, USA}
%\and
%\author{T. del Pino Alem\'an$^{6}$}
%\affiliation{Instituto de Astrof\'isica de Canarias, E-38205 La Laguna, Tenerife, Spain}
%\affiliation{Departamento de Astrof\'isica, Universidad de La Laguna, E-38206 La Laguna, Tenerife, Spain}

\author{C. Erba$^{5}$}
%\affiliation{Department of Physics \& Astronomy, East Tennessee State University, Johnson City, TN 37614, USA}
%\and
\author{C.\ P.\ Folsom$^{6}$}
%\affiliation{Tartu Observatory, University of Tartu, Observatooriumi 1, T\~{o}ravere, 61602, Estonia}

\author{K. Gayley$^{7}$}
%\affiliation{Department of Physics \& Astronomy, University of Iowa, 203 Van Allen Hall, Iowa City, IA, 52242, USA}

%\author{R.\ Ignace$^{11}$}
%\affiliation{Department of Physics \& Astronomy, East Tennessee State University, Johnson City, TN 37614, USA}

%\author{Z. Keszthelyi$^{12}$}
%\affiliation{Anton Pannekoek Institute for Astronomy, University of Amsterdam, Science Park 904, 1098 XH, Amsterdam, The Netherlands}

%\author{O. Kochukhov$^{13}$}
%\affiliation{Department of Physics and Astronomy, Uppsala University, Box 516, 75120 Uppsala, Sweden}

\author{Y. Naz\'e $^{8}$}
%\affiliation{GAPHE, Universit\'e de Li\`ege, All\'ee du 6 Ao\^ut 19c (B5C), B-4000 Sart Tilman, Li\`ege, Belgium}
%\and
\author{C. Neiner$^{9}$}
%\affiliation{LESIA, Paris Observatory, PSL University, CNRS, Sorbonne Universit\'e, Univ. Paris Diderot, Sorbonne Paris Cit\'e, 5 place\\ Jules Janssen, 92195 Meudon, France}
%\and
%\author{M. Oksala$^{16}$}
%\affiliation{Department of Physics, California Lutheran University, 60 West Olsen Road 3700, Thousand Oaks, CA, 91360, USA}

\author{V. Petit$^{10}$}
%\affiliation{Department of Physics and Astronomy, University of Delaware, 217 Sharp Lab, Newark, Delaware, 19716, USA}

\author{R. Prinja$^{11}$}
%\aiffiliation{Department of Physics and Astronomy, Gower Street, London WC1E 6BT, UK }

%\author{P. A. Scowen$^{18}$}
%\affiliation{NASA Goddard Space Flight Center, 8800 Greenbelt Rd., Greenbelt, MD 20771}

\author{M. E. Shultz$^{10}$}

\author{N. Sudnik$^{12}$}
%\affiliation{Nicolaus Copernicus Astronomical Centre of the Polish Academy of Sciences, Bartycka 18, 00-716 Warsaw, Poland}

\author{J. S. Vink$^{13}$}
%\affiliation{Armagh Observatory and Planetarium, College Hill, BT61 9DG Armagh, Northern Ireland}
%\and

%\and
\author{G.\ A.\ Wade$^{14}$}

\abstract{
Several space missions and instruments for UV spectropolarimetry are in preparation, such as the proposed NASA MIDEX Polstar project, the proposed ESA M mission Arago, and the Pollux instrument on the future LUVOIR-like NASA flagship mission. In the frame of Polstar, we have studied the capabilities these observatories would offer to gain information on the magnetic and plasma properties of the magnetospheres of hot stars, helping us test the fundamental hypothesis that magnetospheres should %simultaneously 
act to rapidly drain angular momentum, thereby spinning the star down, whilst simultaneously reducing the net mass-loss rate. Both effects are expected to lead to dramatic differences in the evolution of magnetic vs. non-magnetic stars. 
}

%\keywords{Ultraviolet astronomy (1736);
%Ultraviolet telescopes (1743);
%Space telescopes (1547);
%Circumstellar disks (235);
%Early-type emission stars (428);
%Stellar rotation (1629);
%Spectropolarimetry (1973);
%Instruments: Polstar; UV spectropolarimetry; NASA: MIDEX}
\maketitle
%\begin{keywords}

%\end{keywords}

\section{Introduction}\label{sec:intro}
\extrafloats{100}
\subsection{Background}\label{subsec:background}

The hot, massive stars of the upper main sequence are dominant objects in a galaxy. While much less numerous than cooler stars, they exert a wide-ranging influence on galactic structure and stellar ecology. The majority of the periodic table -- all elements up to iron \citep{johnson_science} -- is forged in their cores, with heavier elements being synthesized in their deaths in Type II supernovae, at which point the enriched material is returned to the interstellar medium (ISM). With tens to hundreds of thousands of times the luminosity of the Sun, and much higher temperatures with spectra peaking in the far ultraviolet (FUV), they contribute most of the ISM's ionizing radiation. Their high luminosities launch powerful, radiatively driven winds with mass loss rates thousands to billions of times the Sun's to terminal velocities of thousands of km/s, injecting substantial matter and momentum into the ISM. 

The combination of ionizing radiation and powerful stellar winds hollows out star-forming material, quenching star formation. Upon supernova detonation at the end of a hot star's life, the resulting shock wave contributes a final burst of rapidly moving material that can trigger star formation by initiating the gravitational collapse of nearby molecular clouds. Following the supernova, the stellar remnant begins an extended afterlife as a neutron star or black hole, long-lived objects that -- in the deep cosmic future -- will be {one of the few} inhabitants of the cosmos { \citep{2010ApJ...717..183R, 2016mssf.book..187S,2022MNRAS.511..953C,2022MNRAS.512..216G,1997RvMP...69..337A}}. 

The structure and evolution of single stars are largely functions of the stellar mass, therefore any process that changes the mass will change the evolution of the star. For O-type stars, which have mass-loss rates of around 10$^{-6}~{\rm M_\odot~yr^{-1}}$, main-sequence lifetimes on the order of 10 Myr, and initial masses of around $50~{\rm M_\odot}$, mass loss via stellar winds can result in considerable reductions in mass. The evolution of such stars therefore cannot be understood in isolation from the effects of their wind, since differences in the terminal, pre-supernova state of a star have consequences for the type of supernova as well as the type of remnant that emerges. 

Many massive stars are rapid rotators, with equatorial surface rotational velocities of hundreds of km/s. In the most extreme cases, stars may rotate near their critical or breakup velocities, leading to deformation of their shape from a spherical to an oblate form, bulged around the equator, with much cooler equatorial temperatures due to gravity darkening \citep{1924MNRAS..84..665V}. Even in less extreme cases, rotation leads to mixing, replenishing the nuclear-burning core with fresh material from the envelope, and thereby having a strong effect on the evolution of the star \citep[e.g.][]{2011A&A...530A.115B}. 

Due to the dominant role played by massive stars in terms of mass and energy input via winds, ionizing radiation, and supernovae, understanding the evolution of galaxies requires understanding the evolution of massive stars, which in turn requires that we understand their winds and their rotation along with all phenomena that can modify these key parameters. 

It is clear that there is no such thing in nature as a really `normal' or `standard' massive or hot star. Instead, there is a diversity of special cases, such as classical Be stars, interacting binaries, or magnetic hot stars, which collectively comprise the hot star population. It is only via understanding of the properties of these individual groups that the properties of the massive stars can be properly accounted for in population synthesis models, which in turn are key inputs for models of galactic chemical and structural evolution. 

Main focus of this paper is understanding the properties of the special group of magnetic hot stars. The proposed 
Polstar NASA MIDEX mission will greatly help us achieve this goal, as it is equipped with a 60-cm telescope and a full-Stokes (IQUV) spectropolarimeter divided in 2 channels in the ultraviolet  with the first one providing spectropolarimetry at high spectral resolution of R$\sim$33000 over the 122-200 nm far-UV bandpass and the second one providing spectropolarimetry over the 180-320 nm NUV band with low- to mid-resolution (R$\sim$30 to 250) \citep{2021SPIE11819E..08S}. Thus, Polstar is quite well suited for studying hot stars and their circumstellar environment.

 In the remainder of this section, general background is provided on hot star magnetic fields, magnetospheres, magnetospheric diagnostics, and the effects of magnetospheres on stellar evolution, culminating with the motivation for exploring these effects with Polstar. The known properties of ultraviolet magnetospheric diagnostics, a comparison with visible and X-ray diagnostics, and an overview of the simulations and models used to interpret them, are given in Sect. \ref{sec:uv_spectroscopy}.  Expectations for linear spectropolarimetry arising from scattering in the circumstellar environment, drawing on both models and observations acquired with high-resolution visible instrumentation, are described in Sect. \ref{sec:lin_specpol}. Linearly polarized broadband magnetospheric signatures, again drawing on both models and observations, are given in Sect. \ref{sec:continuum_linpol}.  Enabled science falling under the purview of magnetic and magnetospheric activity in hot stars is described in Sect.~\ref{sec:enabled}.  The  paper is summarized in Sect. \ref{sec:summary}.

\subsection{Massive star magnetism}\label{subsec:magnetism}

Magnetic fields are a crucial factor that leads to drastic modifications in both the mass-loss rates and rotation of hot stars. 
They are found in approximately 10\% of the OBA population \citep{grunhut2017,2017A&A...599A..66S,2019MNRAS.483.3127S}. The magnetic fields of stars with radiative envelopes share similar properties from approximately spectral type A5 to the top of the main sequence \citep{2009ARA&A..47..333D}. They are strong \citep[ranging from hundreds of G to tens of kG;][]{2019MNRAS.490..274S}, they are stable over timescales of at least decades \citep{2018MNRAS.475.5144S}, and they are globally organized and, with few exceptions, geometrically simple \citep[being well-described by tilted dipoles with most of the magnetic energy in low-order poloidal field components;][]{2019A&A...621A..47K}. Magnetohydrodynamic simulations have demonstrated that such fields, once established, can remain stable over evolutionary timescales \citep[e.g.][]{2004Natur.431..819B}, leading to the hypothesis that they are of `fossil' origin i.e.\ remnants of some previous event in the star's life \citep{2009ARA&A..47..333D}.

Moreover, since there is no correlation of magnetic field strengths with rotational or stellar properties, as expected and observed for the dynamo fields of cool stars \citep[e.g.][]{2016MNRAS.457..580F,2017NatAs...1E.184S}, and as there is furthermore no known dynamo mechanism capable of sustaining globally organized magnetic fields in the radiative atmospheres of hot stars, it supports further our belief that these magnetic fields are indeed fossils arising in some earlier phase of the star's life \citep{2009ARA&A..47..333D}, e.g.\ due to amplification of ISM seed fields during formation and/or convective phases during pre-main sequence evolution \citep[e.g.][]{2019A&A...622A..72V}, or in dynamos powered by binary mergers \citep[e.g.][]{
2019Natur.574..211S}. An improved understanding of the formation mechanism of fossil fields therefore offers the promise of advancing our general understanding of the formation or evolution of hot stars. 

While strong fields are relatively rare in more massive OBA stars, weak fields may be widespread. The detection of `ultra-weak' ($\sim$0.1-10 G) magnetic fields in a number of bright A-type stars suggests that magnetic phenomena may be ubiquitous on the upper main sequence \citep{2020MNRAS.492.5794B}. Weak magnetic fields have also been detected in numerous blue supergiants, indicating that they may play a role in post-main sequence evolution as well \citep{2017MNRAS.471.1926N,2018MNRAS.475.1521M,2021mobs.confE..47O}. These weak fields are expected to arise from small-scale dynamos powered by embedded helium and iron opacity-bump convection zones \citep{2020ApJ...900..113J}, which are revealed as the stellar wind strips away the outer layers of the star \citep{2021arXiv211003695J}. However, in more massive OB stars, even if weak magnetic field incidence were to be found ubiquitous, their influence is expected to be dynamically negligible due to a much stronger wind or equivalently to much higher mass loss rates. 

\subsection{Massive star magnetospheres}\label{subsec:magnetospheres}

Strong magnetic fields are easily capable of trapping the ionized wind of a massive star \citep{udDoula2002}. The surface mass flux from opposite magnetic colatitudes is channeled by the magnetic field and collides in the magnetic equatorial plane, leading to the accumulation of a torus of high-density plasma surrounding the magnetic equator, with X-rays produced in the cooling shocks \citep{2014MNRAS.441.3600U}. 

Magnetic confinement extends out to the Alfv\'en surface, the distance from the star at which the magnetic energy density and the kinetic energy density of the wind equalize, with the Alfv\'en radius $R_{\rm A}$ defined in the magnetic equatorial plane. Alfv\'en radii commonly range from a few stellar radii (as is typical for an O-type star, where the powerful wind rapidly overpowers the magnetic field) to tens of stellar radii (as is typical for a strong magnetic field trapping the much weaker wind of a B-type star). 

Beyond the Alfv\'en surface, the wind is magnetically unconfined and escapes from the star. Below the Alfv\'en surface, plasma is forced into corotation with the stellar magnetic field by the Lorentz force. If a star is slowly rotating, gravity pulls the confined dense plasma back to the star. This can dramatically reduce the net mass-loss rate of the star \citep{udDoula2002,2017MNRAS.466.1052P}. If a star is rapidly rotating, such that the Kepler corotation radius $R_{\rm K}$ (at which gravity is balanced by the centrifugal force imparted by corotation) is inside the Alfv\'en radius, gravitational infall is prevented in the part of the magnetosphere above $R_{\rm K}$. The former case is referred to as a {\em Dynamical Magntosphere} (DM) since mass-balancing occurs on dynamical timescales; the latter case is referred to as a {\em Centrifugal Magnetosphere} \citep[CM;][]{petit2013}. Importantly, all magnetic stars have DMs in at least the inner part of the magnetosphere\footnote{Unless the star is critically rotating, in which case $R_{\rm K}$ becomes identical to the equatorial radius; while theoretically possible, no such object has been found, and in any case such a phase would be extremely short-lived due to strong magnetic braking.}.

Plasma inside a CM is unable to fall back to the star, and 2D magnetohydrodynamic (MHD) simulations indicate that it will instead build up until it reaches a critical density beyond which the magnetic field is unable to confine it, at which point it is  ejected outwards -- a phenomenon referred to as centrifugal breakout \citep{udDoula2008}.  However, more recent 3D MHD simulations indicate such breakout events occur continuously and on a smaller scale  \citep{2020pase.conf..140U}. This mass-balancing mechanism has been verified via analysis of the H$\alpha$ emission properties of stars with CMs \citep{2020MNRAS.499.5379S,2020MNRAS.499.5366O}. Thus, plasma accumulated in the DM is trapped and returned to the star, whereas plasma trapped in the CM will eventually be lost; only the CM reduces the net mass-loss rate. 

Poynting stresses in the magnetosphere lead to rapid angular momentum loss \citep{2009MNRAS.392.1022U}, as a result of which magnetic hot stars are systematically more slowly rotating than non-magnetic stars of comparable spectral type and luminosity class \citep{2018MNRAS.475.5144S}. In some cases, the rotational period change can be directly measured \citep{2010ApJ...714L.318T}. Rotational spindown can be so extreme that rotational periods of up to decades in length have been identified \citep[e.g.][]{2010A&A...520A..59N,2017MNRAS.468.3985S,2017MNRAS.471.2286S,2021MNRAS.506.2296E}.  An important consequence of this is that the CM quickly shrinks as a star ages, and is detectable only during the initial phase of the main sequence \citep{2019MNRAS.490..274S}.

\subsection{Multiwavelength magnetospheric diagnostics}\label{subsec:diagnostics}

In addition to their important consequences for stellar evolution, magnetospheres have a number of observational consequences. There are available diagnostics across the electromagnetic spectrum, each of which probes a different magnetospheric component, and is detectable in a different part of stellar and magnetospheric parameter space. 

X-rays are emitted due to magnetically confined wind shocks, making magnetic stars much more luminous in X-rays than non-magnetic stars \citep[e.g.][]{2014ApJS..215...10N,2014MNRAS.441.3600U}. While X-rays are detectable for most magnetic OB stars, time series data are difficult to acquire. 

Velocity-resolved information from line emission is furthermore difficult to obtain for all but the brightest X-ray sources. However, in 2031 the launch of the {\em Athena} mission will in fact provide velocity-resolved X-ray information \citep{2013arXiv1306.2307N}, which may provide a powerful complement to the data gathered by Polstar (although the science case presented here is independent of any results from Athena). 

Rapidly rotating magnetic AB stars\footnote{This diagnostic is unavailable for O-type stars, since the large radio photospheres produced by their dense winds swallow any gyrosynchrotron radiation produced by their magnetospheres \citep[e.g.][]{2015MNRAS.452.1245C}.} can exhibit radio gyrosynchrotron emission originating at high magnetic latitudes \citep[e.g.][]{2021MNRAS.507.1979L}, beamed auroral radio emission via the electron cyclotron mechanism \citep[e.g.][]{2021arXiv210904043D}, and line emission in near infrared and visible Bracket, Paschen, and Balmer series H lines, most prominently in H$\alpha$ \citep[e.g.][]{grun2012,2015A&A...578A.112O}. 

Since the weak winds of B-type stars are unable to fill their DMs to sufficient density to become optically thick in H$\alpha$ \citep{petit2013}, their magnetospheres are detectable in visible data only around young, rapidly rotating, strongly magnetic stars with large CMs, which due to rapid magnetic braking disappear after less than 1/3 of the stars' main sequence lifetime \citep{2019MNRAS.490..274S}. 

By contrast, ultraviolet signatures associated with magnetospheres have been detected across the Hertzsprung-Russell diagram, 
making FUV resonance lines the most versatile and ubiquitous diagnostic by far. These lines probe both the free, unconfined stellar wind, and the dynamical magnetosphere \citep[e.g.][]{1990ApJ...365..665S,mar13,petit2013,erb21}. However, the C IV wind lines persist to B3 or so, and below that wind signatures disappear in resonance lines \citep{1994ApJS...94..163S}. While \citet{2015MNRAS.451.2015O} showed that modelling of the equivalent width of SiIV line could plausibly be reproduced with a photospheric model, they did not
give detailed line profile models. But, in general, UV diagnostics provide information on that part of the wind which is escaping, and that part which is returned, to the star. UV resonance line variability is almost universally detectable in magnetic OB stars \citep[e.g.][]{2001ASPC..248..393H,2008A&A...483..857S,2012A&A...545A.119H,2013A&A...555A..46H,2019MNRAS.483.2814D,2021MNRAS.506.2296E}. It therefore provides the only diagnostic that a) can be used for magnetic stars across the full range of magnetic, rotational, and stellar parameters, b) provides simultaneous information on the wind and the magnetosphere, and c) provides velocity-resolved information with which the detailed magnetospheric structure can be examined. 

The application of UV spectroscopy to the study of hot star magnetospheres is conceptually similar to its utility in the study of the solar magnetosphere. One recent example is the IRIS mission, which has been obtaining high resolution FUV spectroscopy of solar plasmas since 2013. The FUV band of IRIS regularly observes the same Si~{\sc IV} doublet that is used to probe the winds of hot stars. In the solar case where Si~{\sc iv} line profiles are in pure emission, these profiles  are used to probe Doppler flows of plasmas at $T \sim 80$~kK \citep[e.g.][]{2015ApJ...801...83C}.

\section{UV spectroscopic diagnostics of massive star magnetospheres}\label{sec:uv_spectroscopy}

   \begin{figure}[t]
   \centering
   \includegraphics[width=0.48\textwidth,trim = 50 50 50 0]{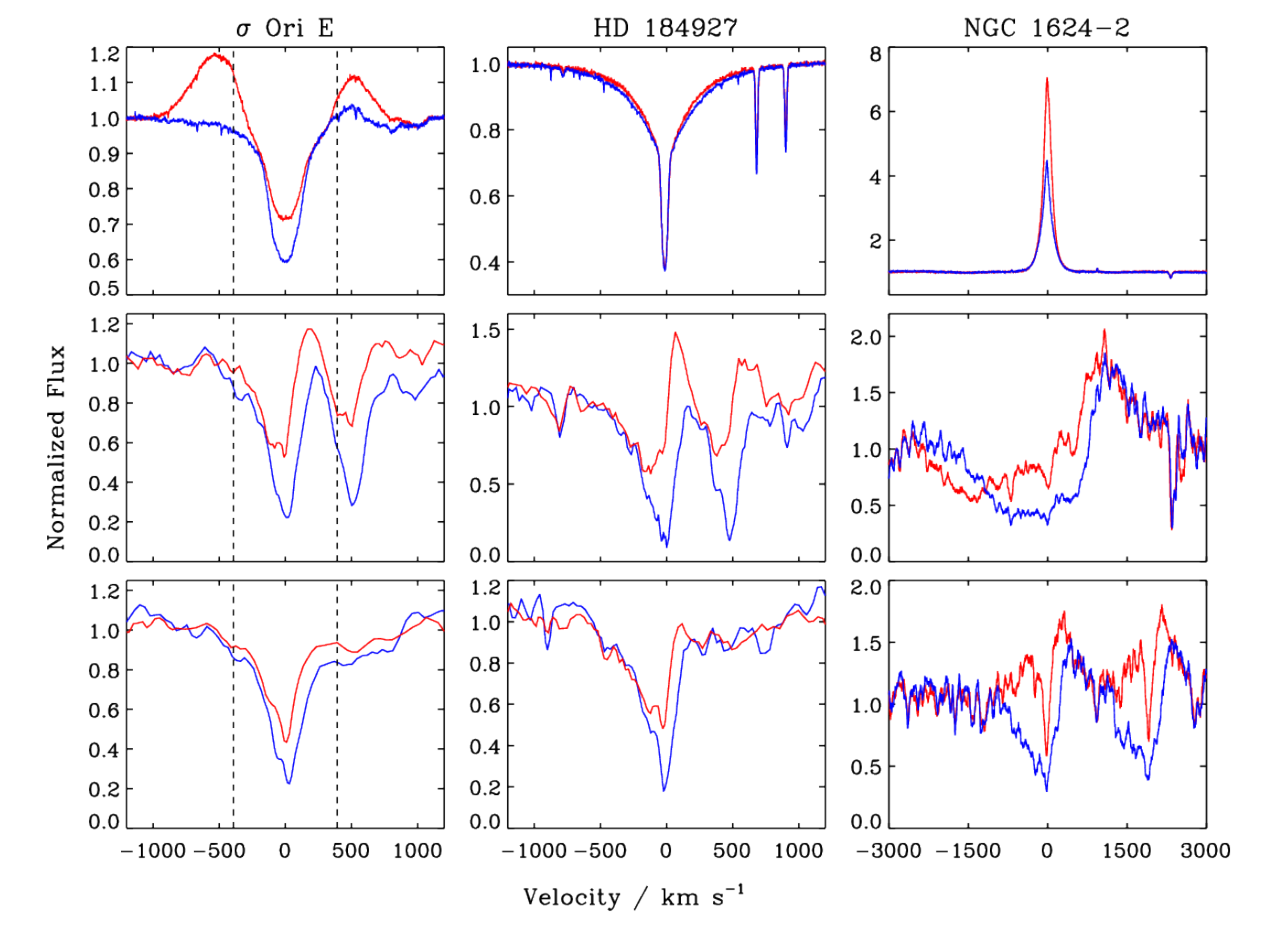}
      \caption{Comparison of H$\alpha$ (top), the C~{\sc iv} $\lambda\lambda$1548,1551 doublet (middle), and the Si~{\sc iv} $\lambda\lambda$1394,1403 doublet (bottom) for 3 representative stars (note that the red component of the Si~{\sc iv} is not visible for $\sigma$ Ori E and HD\,184927). High and low state are respectively indicated by red and blue. For $\sigma$ Ori E, dashed lines indicate $\pm R_{\rm K}$. All lines have been shifted to the stellar rest frame. For more details, see for instance \citet{2012MNRAS.425.1278W}}
         \label{polstar_uv_halpha}
   \end{figure}

The interest of UV observations has been well demonstrated in the study of magnetic O stars \citep[e.g.][]{smi05,gru09,mar12,mar13,naz15,2019MNRAS.483.2814D,dav21}. As UV line profiles trace the density and velocity structure of the wind along the line-of-sight, line profile variations are expected for magnetic stars as the stellar rotation modifies the observer's viewing angle of the non-spherical magnetosphere. The available observations reveal enhanced absorption at low-velocity where the slow, confined wind appears in front of the star (equator-on view). In contrast, high-velocity absorption is recorded for optically-thick lines where the fast, polar wind enters the line-of-sight (pole-on view). These differences between pole-on and equator-on observations are magnified for extremely magnetic stars (such as NGC 1624-2), or are totally damped in the case of very weakly magnetic stars (e.g., $\zeta$ Ori A). However, they are qualitatively reproduced by dedicated models that in general assume simple dipolar magnetic field topology \citep{mar13,naz15,erb21} indicating a smooth transition between these two phases. 

But additional observational datasets obtained at different stellar rotational phases reveal clear departures from such a smooth transition \citep{dav21}.  Furthermore, for some  cases,  the UV lines profiles along opposite poles exhibit significant differences \citep{naz15}, indicating a more complex magnetospheric geometry than the generally assumed simple dipoles.  By observing a large sample of stars with a dense phase coverage, Polstar will reveal the  magnetic topology more accurately. 

UV spectroscopy has also been crucial to detecting and understanding the magnetospheres of B-type stars \citep[e.g.][]{2001A&A...372..208S,2003A&A...406.1019N,2003A&A...411..565N,2006MNRAS.370..629D,2008A&A...483..857S,2011MNRAS.412L..45P,2013A&A...555A..46H,2021MNRAS.506.2296E}, including the ones lacking the surface chemical peculiarities characterizing the universally magnetic Bp class \citep[e.g.][]{2003A&A...406.1019N,2006MNRAS.370..629D,2011MNRAS.412L..45P,2017MNRAS.471.2286S}.  As surface chemical abundance patches are unable to form when the stellar wind strips the surface
faster than patches can accumulate, these anomalous ultraviolet lines enabled successful identification
of magnetic stars earlier than B2, as in the case of $\beta$~Cep \citep{2013A&A...555A..46H}, $\tau$~Sco \citep{Donati2006-ZDI-SHD}, or the $\tau$~Sco analogues \citep{2011MNRAS.412L..45P}.

As with magnetic O-type stars, magnetic B-type stars exhibit periodic variations modulated on the rotational timescale. Observationally, the UV lines of magnetic B stars are characterized by a `high state' with a red-shifted emission feature and a blue-shifted absorption, similar to a the classical P-Cygni profile originating in a spherically expanding wind, but appearing at much cooler effective temperatures than those at which P-Cygni profiles are generally seen. At `low state', the red-shifted emission disappears and the blue-shifted emission becomes stronger. The high state corresponds to the maxima of \bz, the line-of-sight magnetic field averaged over the visible stellar hemisphere, i.e.\ high state is seen when the magnetic pole is closest to the line of sight. On the other hand, low state corresponds to \bz~nulls, i.e.\ when the magnetic equator is along the line of sight. 
In terms of modeling, this is generally interpreted as a result of a plasma torus in the magnetic equatorial plane, which either produces emission when viewed pole-on, or absorption when the torus eclipses the star in equator-on view.
%together with an absorption component produced by the escaping wind itself. 

Magnetic B-type stars frequently display variability in the N~{\sc v} 123.9, 124.3 nm doublet. These lines are not easily produced in such stars since the ionization temperature of N~{\sc iv} is higher than their photospheric effective temperature. This is believed to be due to over-ionization from X-rays produced in colliding wind shocks \citep[e.g.,][]{2001A&A...372..208S,2014MNRAS.441.3600U}, and is entirely absent in non-magnetic B-type stars.
As such, this doublet can provide us with a clear diagnostics for magnetism in B-type stars.  NLTE calculations, like those done in another context by \citet{2016A&A...590A..88C}, will be required before a full understanding is achieved.

\subsection{Visible versus Ultraviolet Magnetospheric Diagnostics}\label{subsec:vis_uv_spectroscopy}

Fig.\ \ref{polstar_uv_halpha} compares the variable magnetospheric line diagnostics available from H$\alpha$ to two UV resonance lines, the C~{\sc iv} 154.8, 155.1 nm and Si~{\sc iv} 139.4, 140.3 nm doublets. 

In the case of the CM star $\sigma$ Ori E \citep[B2\,Vp;][]{1977PASP...89..797A}, H$\alpha$ emission variability is predominantly located outside of the Kepler corotation radius (indicated with vertical dashed lines), a consequence of the accumulation of cool, dense plasma at and beyond $R_{\rm K}$ which, at high state, is projected off of the stellar limb. By contrast, neither of the UV doublets demonstrate any variable emission outside of $\pm R_{\rm K}$. Instead, there is an emission feature shifted to the red of the line center. At low state, the high-velocity emission disappears in H$\alpha$, replaced by enhanced absorption in the line core; this is due to the CM eclipsing the star \citep[e.g.,][]{2012MNRAS.423L..21S, udDoula2013}. 
Similar enhanced absorption, also due to eclipsing, is detectable in the UV during low state. Note that the amplitude of variation is largest in the C~{\sc iv} doublet, however as the two components are very close to one another their emission overlaps, whereas the Si~{\sc iv} doublet has weaker emission, but the larger separation of its components enables the contribution of each line to be more effectively isolated. 

There is no H$\alpha$ emission in the case of HD\,184927 \citep[B2\,Vp;][]{2015MNRAS.447.1418Y}: while this star has a similarly strong magnetic field to that of $\sigma$ Ori E, it has a much longer rotational period and, therefore, does not possess a large CM. Its UV variability is however remarkably similar to that of $\sigma$ Ori E. Together with the absence of UV variations outside the Kepler radius in $\sigma$ Ori E's UV lines, this suggests that, amongst B-type stars, the UV is primarily probing the dynamical part of the magnetosphere, i.e.\ the region closest to the star \citep[e.g.][]{petit2013}. 

The O-type star NGC\,1624-2 has the strongest magnetic field of any star at the top of the main sequence \citep[$B_{\rm p}~\sim20$~kG;][]{2012MNRAS.425.1278W}. Its long rotational period ($\sim158$~d) 
means that its magnetosphere is purely dynamical \citep{petit2013}. Its H$\alpha$ line shows extremely strong emission at high state, due to its large Alfv\'en radius and the powerful wind that easily fills the DM. H$\alpha$ emission is only slightly weaker during low state. The emission is confined within about $\pm 200$~\kms\ of line centre. The velocity broadening is expected to be primarily thermal and turbulent, and the low velocity of the material again reflects the fact that H$\alpha$ is probing the dense, cool plasma around the magnetic equator \citep[e.g.,][]{2012MNRAS.423L..21S}. By contrast, NGC\,1624-2's UV profiles demonstrate both strong emission and absorption \citep{2019MNRAS.483.2814D,dav21}. The C~{\sc iv} doublet provides a sensitive probe of the unconfined wind, as revealed by its P Cygni-like profile, whereas Si~{\sc iv} is apparently more sensitive to the magnetically confined plasma. 

Fig.\ \ref{polstar_uv_halpha} demonstrates two important advantages of the UV: 1) unlike H$\alpha$, UV emission is detectable for essentially all magnetic stars earlier than B2; 2) even when H$\alpha$ is available, UV emission probes magnetospheric regions that are not reachable by other means. The relevance of UV diagnostics to the characterization of hot star magnetospheres underscores the importance of innovative UV observatories, such as the proposed Polstar mission.

\subsection{X-ray versus Ultraviolet Magnetospheric Diagnostics}\label{subsec:xray_uv}

Relatively strong magnetic fields channel wind towards the magnetic equator where they collide. The resultant strong shock is able to generate hard X-ray emission, in addition to the soft X-rays naturally generated in all magnetic stars earlier than about B3 winds due to the Line-Deshadowing Instability (LDI) \citep{1997A&A...323..121B,1988ApJ...335..914O,2021arXiv211005302D}. For example, in non-magnetic O-stars, LDIs generate soft X-ray emission with a luminosity scaling with the bolometric one ($L_{\rm X}/L_{\rm BOL}\sim 10^{-7}$, \citealt{1996A&AS..118..481B,1997A&A...322..878F}) while the magnetic O-stars show a clear enhancement, (with $\log [L_{\rm X}/L_{\rm BOL}]\sim-6.2$; \citealt{2014ApJS..215...10N,2011MNRAS.416.1456O}). For most B-stars, the X-ray luminosity and its ratio with respect to bolometric luminosity are smaller. However, as for O-stars, they agree well with the predictions of the X-ray Analytic Dynamical Magnetosphere (XADM) model (\citealt{2014ApJS..215...10N, Owocki2016, 2018CoSka..48..144F}). However, a few discrepancies remain and require data at other wavelengths -- especially in the UV -- to understand the peculiarities of the magnetospheres of these targets (e.g. the complex magnetic geometry of $\tau$~Sco, or the potential impact of fast rotation) and the exact temperature stratification (to understand notably why X-rays are not as hard as predicted see e.g. HD~191612, \citealt{2016ApJ...831..138N}, or to understand notably the diversity in X-ray hardness and variability behaviours  (\citealt{2014ApJS..215...10N}). 

In a few cases (e.g., $\theta^1$~Ori~C, \citealt{2005ApJ...628..986G}%Gagn\'e et al. 2005 
; HD~191612, \citealt{2010A&A...520A..59N}
; CPD~--28 2561, \citealt{naz15}%Naz\'e et al. 2015
; NGC~1624-2, \citealt{2015MNRAS.453.3288P}%Petit et al. 2015
), sufficient X-ray data were collected to detect variability of the X-ray flux with the rotational period. Only the extremely magnetic NGC~1624-2 showed a clear increase of absorption when the confined winds were seen magnetic equator-on, demonstrating that the X-ray emitting source lies in the confined winds, and that their cool component can be dense enough to absorb X-rays.  Other objects, in  general, are subject to  occultation effects wherein the X-ray emission directly  behind the star is hidden as the confined winds are viewed magnetic equator-on. In some cases, large amount of X-ray variability  suggests that the X-ray emitting region is not symmetric, as in a ring \citep{2016AdSpR..58..680U}. %(ud-Doula \& Naz\'e 2016) 
However, additional information potentially due to UV polarimetry from Polstar will enable determination of the exact magnetospheric geometry and help us model them more precisely, leading to better constraints in X-rays as well.

Very few magnetic stars can be studied at high-resolution in X-rays because of the low sensitivity of such facilities (although this may change when Athena becomes operative). The narrow and symmetric X-ray lines agree with predictions of magnetohydrodynamic (MHD) models \citep{2003A&A...398..203M,2005ApJ...628..986G,2007MNRAS.375..145N,2008AJ....135.1946N, 2012ApJ...746..142N,2009A&A...495..217F}, %(e.g. Mewe et al. 2003, Gagn\'e et al. 2005, Naz\'e et al. 2007,2008,2012, Favata et al. 2009, vs Asif for theory?) 
but a detailed, quantitative assessment of the plasma kinematics can only be done at other wavelengths. A careful and systematic study of these magnetospheres in UV with  missions such as proposed Polstar  represents a unique opportunity to better understand these objects.

\subsection{Numerical Modeling of Hot Star Magnetospheres}\label{subsec:numerical_models}

\begin{figure*}[t]
%\begin{center}
\vfill
\includegraphics[width=0.33\textwidth]{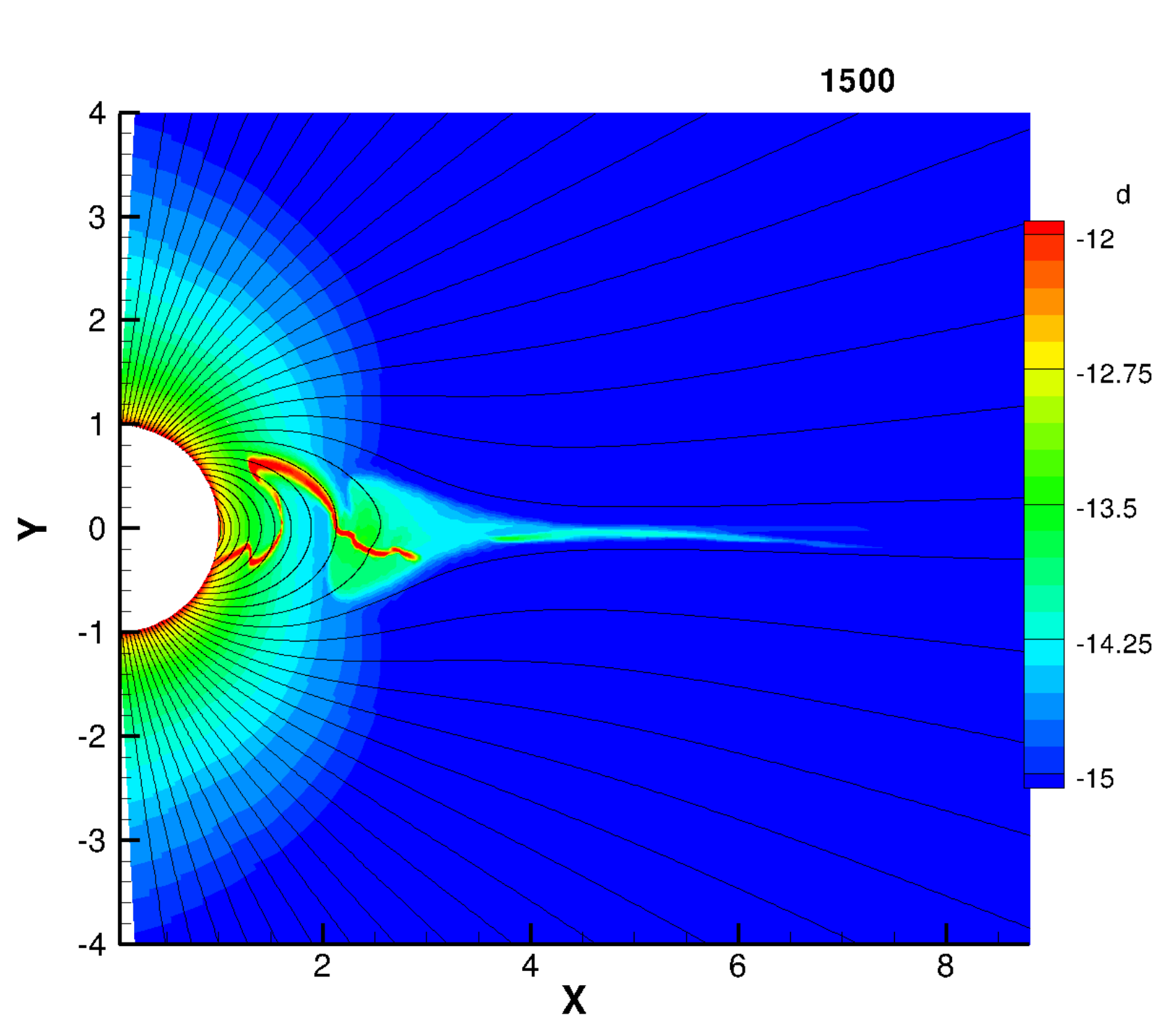}
\includegraphics[width=0.33\textwidth]{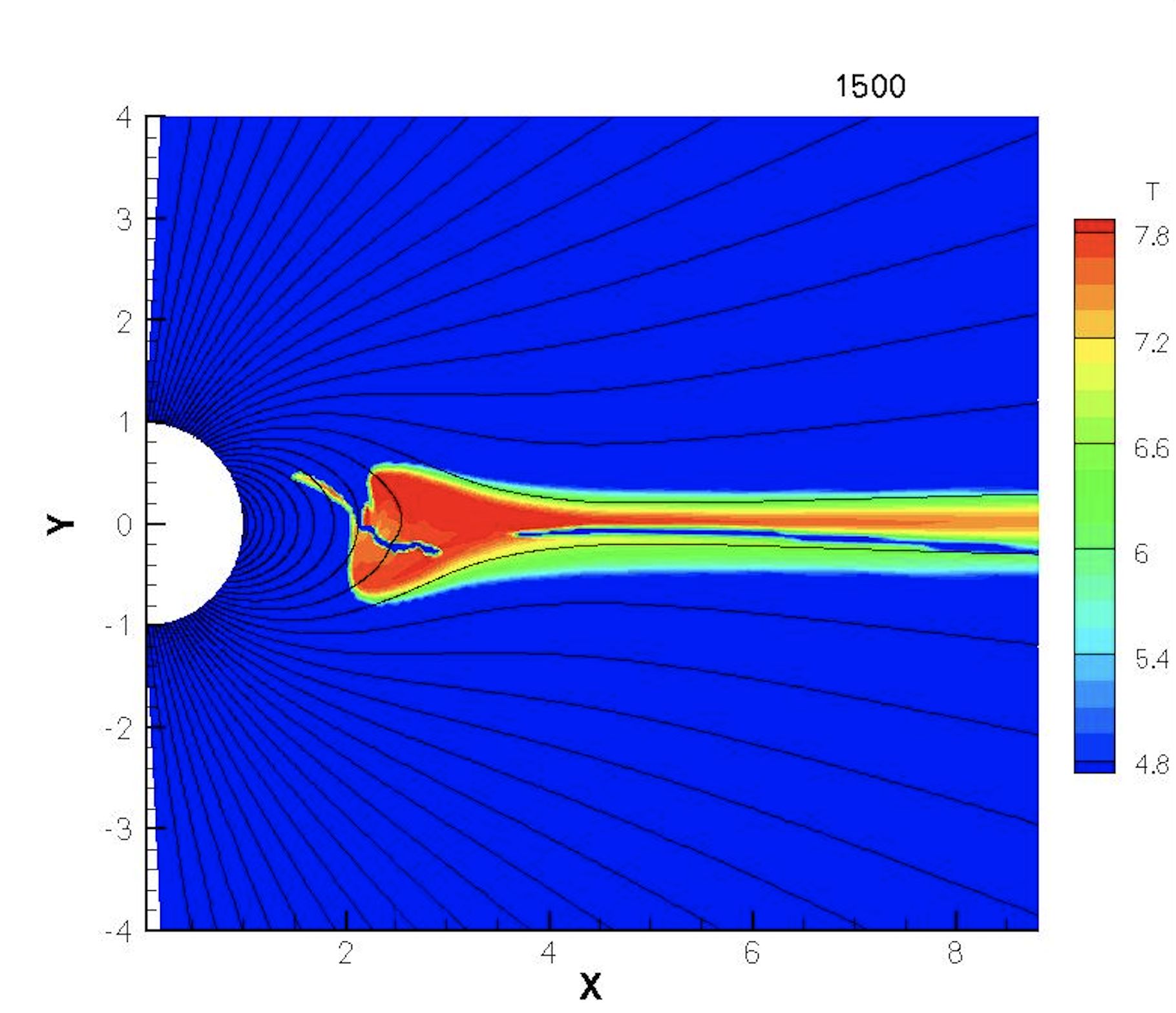}
\includegraphics[width=0.33\textwidth]{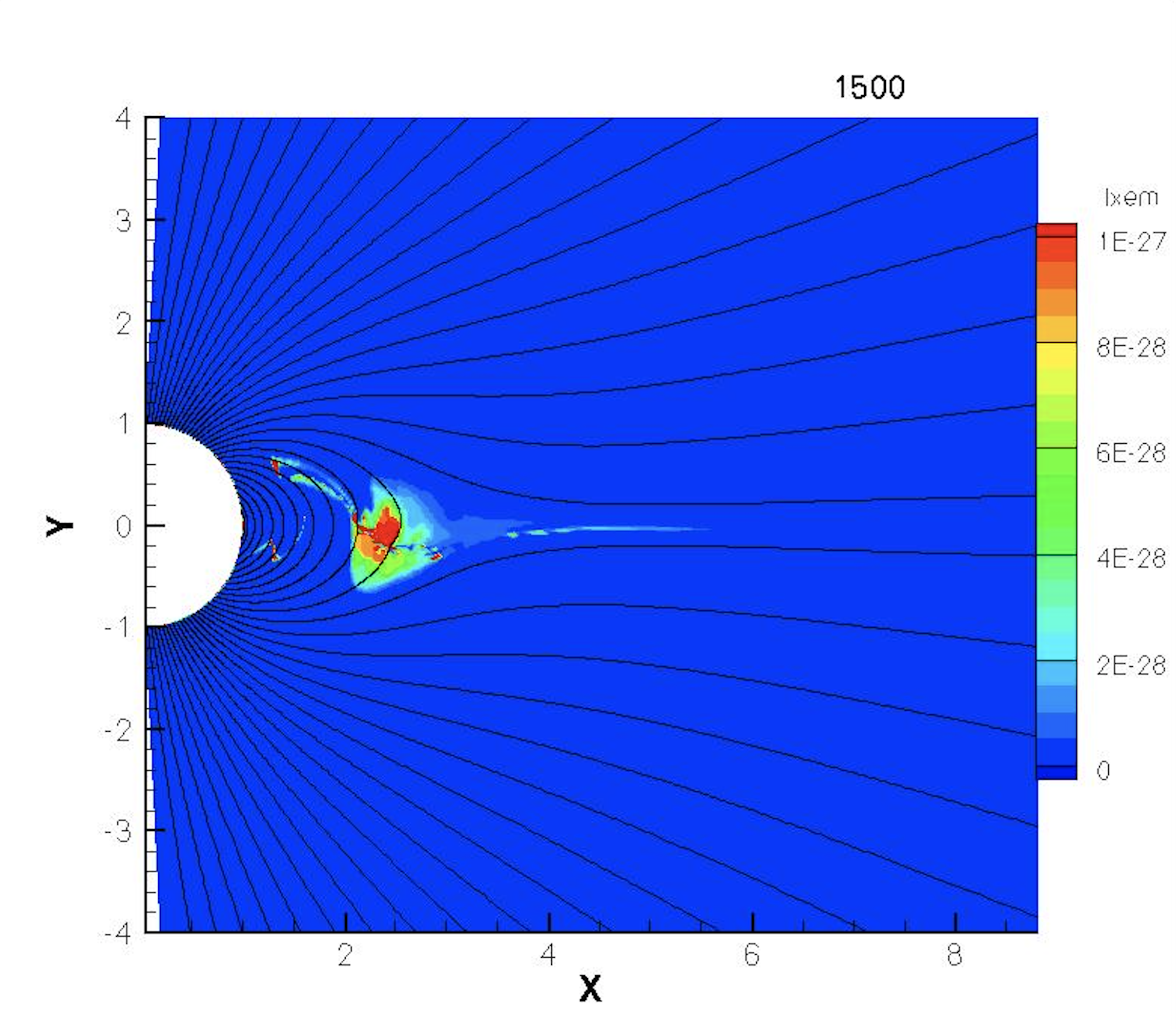}
\caption{
For a 2D MHD simulation of a magnetized wind with confinement parameter $\eta_\ast = 100$,
 color plots of log density (left) and log temperature (middle) in cgs units for arbitrary snapshot many dynamical times after initialization.
 Note that magnetic loops extending above $R_A/R_\ast \approx 100^{1/4} \approx 3.2$ are drawn open by the wind, while those with an apex below $R_A$ remain closed.
The right panel plots associated X-ray emission from the magnetically confined wind shock (MCWS) near the apex of closed loops. Figure  adopted from \citet{udDoula2014}.
}
\label{fig:rhotxem}
%\end{center}
\end{figure*}

\begin{figure*}[t]
%\begin{center}
%	\includegraphics[scale=0.58]{stanfigs/owocki-stan_fig2.pdf}
	\includegraphics[width=0.95\textwidth]{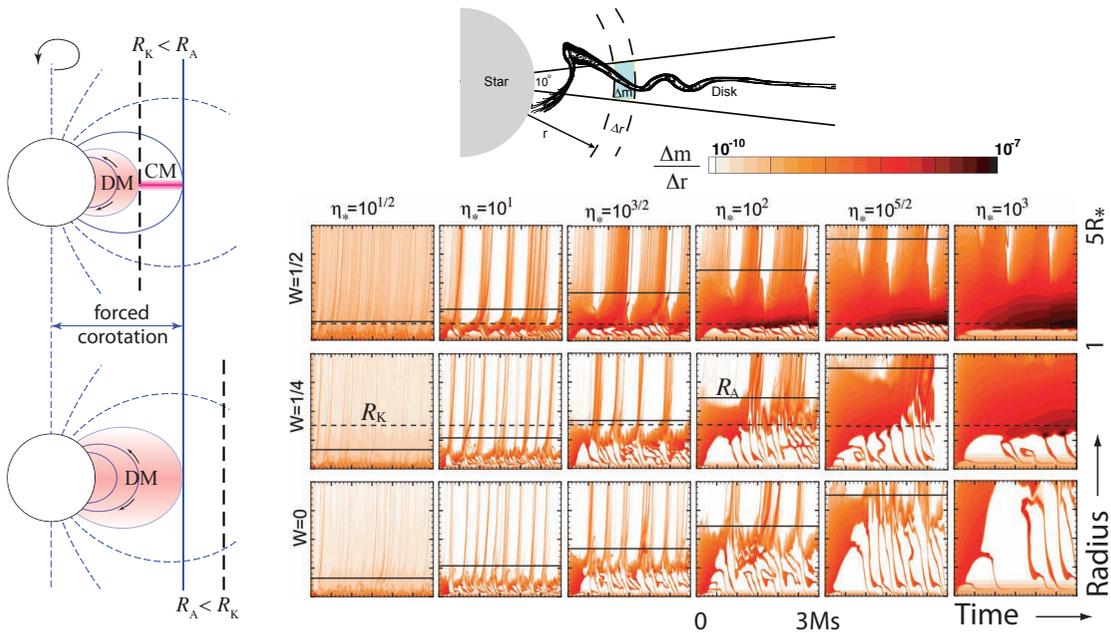}
\vspace{-0.7in}
\caption{
{\em Left:}
Sketch of the regimes for a dynamical vs.\ centrifugal magnetosphere (DM vs. CM).
{\em Upper right:}
 Contour plot for density at an arbitrary snapshot of an isothermal 2D MHD simulation, overlaid with illustration to define the logarithm of radial mass distribution, $\Delta m/\Delta r$ near the equator in the unit of solar mass per stellar radius.  
{\em Lower right:}
Color plots for  log of $\Delta m/\Delta r$, plotted versus radius (1-5 $R_\ast$) and time (0-3~Msec), for a mosaic of 2-D MHD models with a wide range of magnetic confinement parameters $\eta_\ast$, and 3 orbital rotation fractions $W$. 
The horizontal solid lines indicate the Alfv\'en radius $R_{Alfven}$ (solid) and Kepler radius $R_{Kepler}$ (dashed). Figure adopted from \citet{udDoula2008}.
\label{fig:dmdr} 
}
%\end{center}
\end{figure*}

\begin{figure*}[t]
	\begin{center}
	\includegraphics[scale=.6]{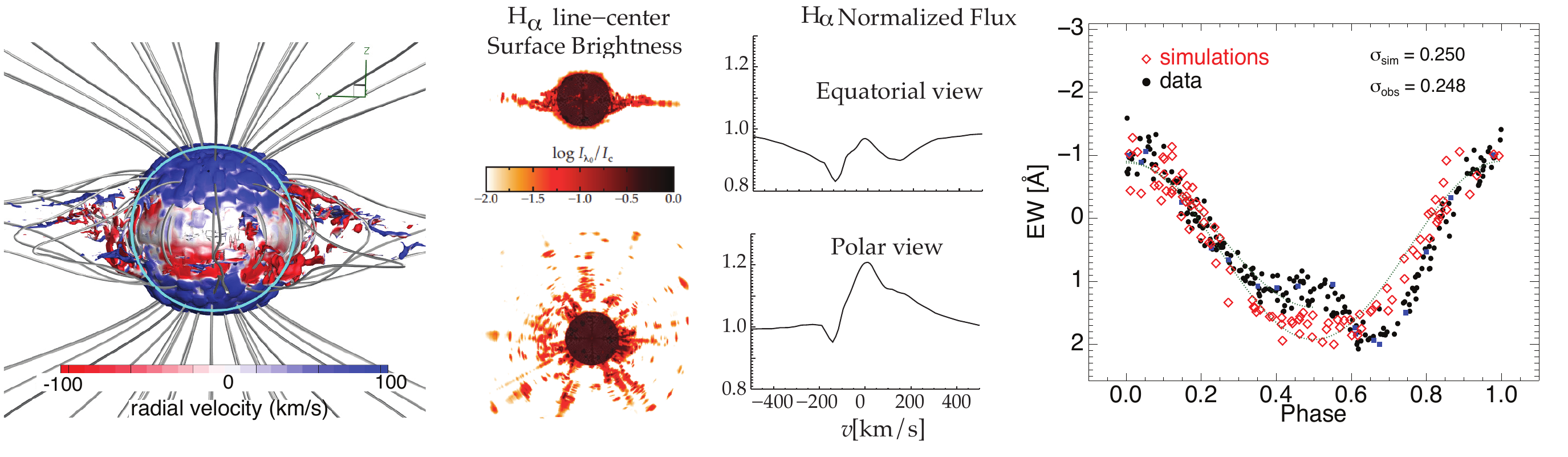}
	\caption{\small{
3D MHD model of the dynamical magnetosphere for the young, slowly rotating (15.4-day period) O7V star $\theta^1$~Ori~C.
%(ud-Doula et al.\ 2013)
The left panel shows a snapshot of wind structure drawn as isodensity surface,
colored to show radial component of velocity.
The middle panels shows the predicted equatorial and polar views of H$\alpha$ line-center surface brightness, along with corresponding line-flux profiles.
The right panel compares the observed rotational modulation of the H$\alpha$ equivalent width (black) with 3D model predictions (red) assuming a pure-dipole surface field tilted by $\beta = 45^\circ$ to the rotation axis, as viewed from the inferred observer inclination of $i = 45^\circ$. Figure adopted from \citet{udDoula2013}.
 }
 }
 \label{fig:3DT1OC}
\end{center}
\end{figure*}

Some characteristics of UV observations can be predicted by numerical models of hot star magnetospheres. In the past two decades, extensive observational surveys supported by theoretical analyses provided a strong basis to  develop a  successful general paradigm for characterizing the properties of hot star magnetospheres in terms of their rotation (setting the Kepler co-rotation radius $R_K$) and level of wind magnetic confinement (setting the Alfv\'en radius $R_A$; see figures \ref{fig:rhotxem} and  \ref{fig:dmdr}).

As with X-rays, MHD simulations have been used advantageously to reproduce the overall variability phenomenology of UV resonance lines \citep{mar13, naz15}. As a sample of a fully self-consistent 3D MHD simulation of a hot star magnetosphere, figure \ref{fig:3DT1OC} \citep{udDoula2013} shows how wind material trapped in closed loops over the magnetic equator in $\theta^1$~Ori~C (left panel) leads to circumstellar emission that is strongest during rotational phases corresponding to pole-on views (middle panel). For a pure dipole with the inferred magnetic tilt  $\beta=45^\circ$, an observer with  the inferred inclination $i=45^\circ$ has  perspectives that vary from magnetic pole to equator, leading in the 3D model to the rotational phase variations in the H$\alpha$ equivalent widths shown in the right panel (shaded circles). Such models can also help us synthesize UV line profiles that Polstar can observe.

Additionally, dynamic flows formed in the DM as material is launched from the surface and then falls back in complex patterns, occurring on timescales of tens of ks \citep{udDoula2008}, could lead to small-scale stochastic variability in several diagnostics. This was tentatively detected for $\theta^1$~Ori~C \citep{udDoula2013}, in which short-timescale variability is found in the equivalent width measurements of H$\alpha$, on top of larger, rotationally-modulated variations. However, such short-term variability has yet to be detected in the UV, as no magnetic star has been observed with sufficiently closely-spaced or high-SNR spectroscopic time-series. To best disentangle the various scales of variability that might arise as a consequence of dynamic flows in the DMs of magnetic massive stars, targets with very dense DMs and very slow rotation \citep[such as HD 108;][]{2010MNRAS.407.1423M,2017MNRAS.468.3985S} appear most promising.

\subsection{Analytical Models of Highly Magnetized Hot Star Winds}\label{subsec:analytic_models}

\begin{figure}
%\centering
%\includegraphics[width=0.45\textwidth]{asifS1_theta1OriC_UVADM.pdf}
\makebox[\columnwidth][c]{\includegraphics[width=\columnwidth]{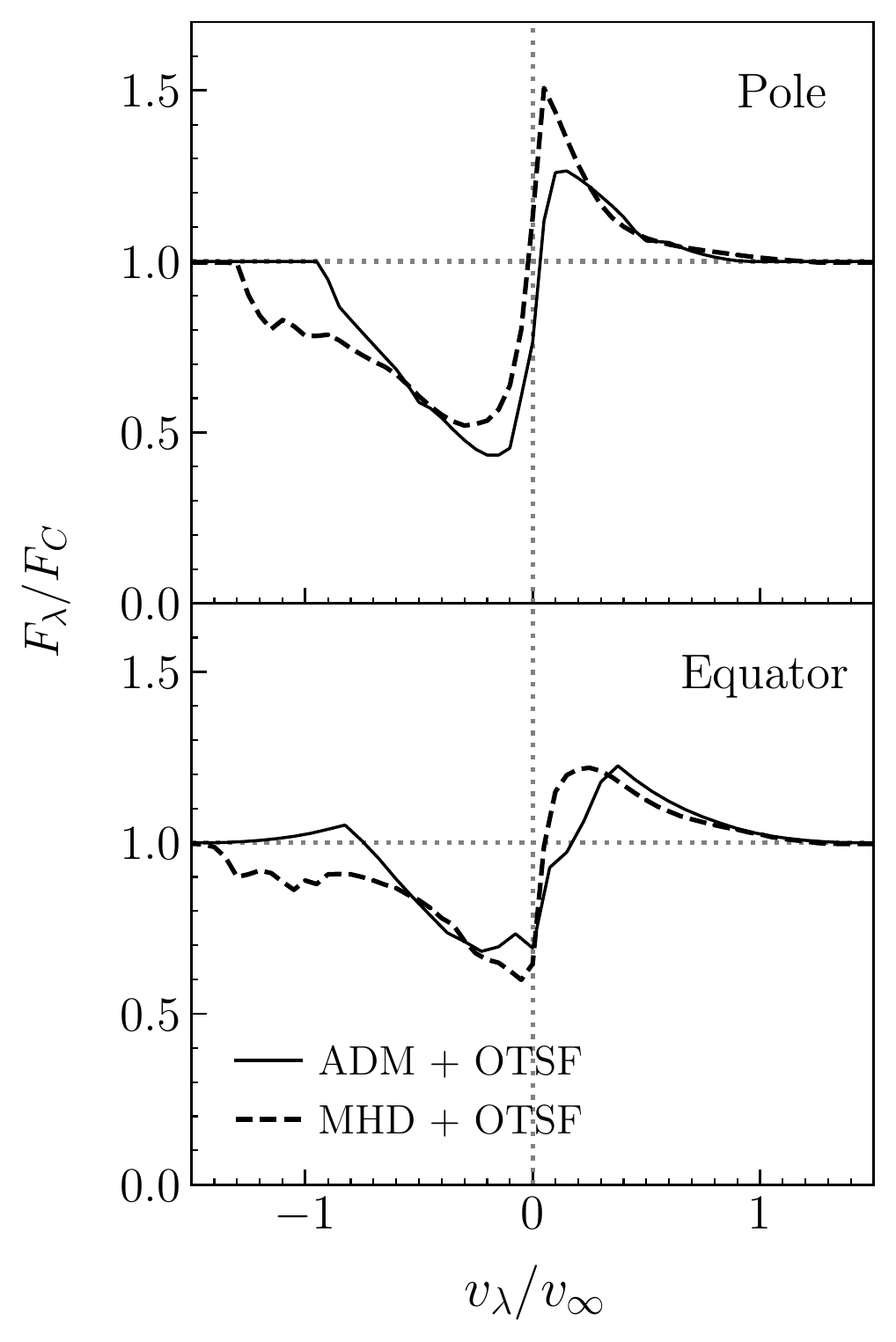}}
\caption{Two sets of synthetic UV line profiles are compared for a magnetic pole-on (upper panel) and equator-on (lower panel) view. The first set of models (solid lines) calculates the density and velocity structure of the magnetosphere using the ADM formalism \citep{erb21} and model parameters similar to the star $\theta^1$~Ori~C ($R_{\rm A} = 2.3~R_*, \; v_\infty = 3200$~\kms) ; the second set (dashed lines) is calculated using a snapshot of the 3D MHD magnetosphere generated for $\theta^1$~Ori~C \citep{udDoula2013}. To produce the line profiles, both sets of models are coupled with the same radiative transfer method that employs the optically thin source function (OTSF). Overall, the line profiles from these two model sets have similar characteristic shapes.
%, demonstrating the line profiles produced using the ADM model agree well with the predictions from MHD.
}
\label{fig:mhd_uvadm_compare}
\end{figure}

Ultimately, proper self-consistent treatment of the global mass budget, including complex cycles of upflow and infall in DMs, and eventual {\em centrifugal breakout} (CBO) of trapped plasma in CMs, requires self-consistent MHD simulations that account for the competition between field and material flow. However, in the limit of arbitrarily strong fields (effectively with $R_A \rightarrow \infty$), analyses based on an idealization of purely rigid fields have led to both the {\em Analytic Dynamical Magnetosphere} (ADM) and {\em Rigidly Rotating Magnetosphere} (RRM) models that have shown great promise for analyzing broad observational trends in multiple diagnostics, without the computational complexity and expense of full MHD simulations. 

\citet{2005ApJ...630L..81T} have shown that the RRM model can successfully reproduce the periodic modulations observed in the light curve, H$\alpha$ emission-line profile, and longitudinal field strength of $\sigma$ Ori E, a rapidly rotating, strongly magnetized star with a large CM. In this model, plasma accumulates at minima within the gravitocentrifugal potential of a rotating star, which define a warped accumulation surface approximately in the magnetic equatorial plane, with the densest regions at the intersections of this plane with the rotational equatorial plane, and an inner edge approximately defined by the Kepler corotation radius $R_{\rm K}$ (with a gap between $R_{\rm K}$ and the star). The model can be extended from a simple tilted dipole to any arbitrary magnetic geometry, e.g.\ as determined via ZDI, a method which \cite{2015MNRAS.451.2015O} used to reproduce the spectroscopic and photometric properties of the prototypical magnetic star $\sigma$ Ori E. The RRM model has also been used to predict the broadband polarimetric observables of $\sigma$ Ori E, although in this case there is severe tension between the best fit achieved by polarimetry and photometry \citep[][see also Sect.\ \ref{sec:continuum_linpol}]{2013ApJ...766L...9C}.

The ADM model \citep{Owocki2016} was developed to provide an approximate, static view of the magnetospheric structure of a slowly-rotating massive star. Compared to time-averaged MHD simulations, this analytic prescription provides a satisfactory description of magnetospheres, and can be used to synthesize various diagnostics (e.g. H$\alpha$ and optical photometric measurements; \citealt{Owocki2016,munoz20}), as well as the strong, wind-sensitive lines found in the ultraviolet \citep{Hennicker2018,erb21}. Figure \ref{fig:mhd_uvadm_compare} shows that when coupled with appropriate radiative transfer techniques, UV line profiles synthesized using the ADM formalism generally compare well with those produced using an MHD magnetosphere.

The ADM model relies on a few simplifying assumptions: rotation is not taken into account, the dynamic flows formed by alternating episodes of wind launching and infall back onto the stellar surface are approximated by using an unphysical superposition of an upflow and a downflow component, and the magnetic field is idealized as a pure dipole. The latter assumption can be relaxed to take into account various values of $R_\textrm{A}$ (e.g. \citealt{erb21}), and even an arbitrary magnetic geometry \citep{2017IAUS..329..369F}.

Given its success in reproducing a range of observational diagnostics, the ADM model represents a computationally inexpensive alternative to MHD simulations, especially given the wide parameter space probed in UV studies of magnetospheres. \citet{erb21} developed a grid of synthetic UV line profiles using the ADM formalism coupled with a simplified radiative transfer technique (using their \textit{UV-ADM} code). They reported the first large-scale parameter study of several factors (e.g. magnetosphere size, line strength, observer's viewing angle) that affect UV wind line formation, with model parameters chosen to correspond with specific magnetic massive stars (e.g. HD 191612, NGC 1624-2). 

The application of the ADM model to UV spectral line synthesis by \citet{erb21} has also provided evidence for spectral features in the ultraviolet that appear to be unique to magnetic massive stars, including the presence of red-shifted absorption and a strong desaturation of the high-velocity absorption trough, resulting in a spectrum which appears to be of a later type than would be assumed from other observations (e.g. optical). Such signatures could help establish ultraviolet spectroscopy as a means of indirectly detecting magnetic fields in massive stars, similarly to other rotationally-modulated variations across the electromagnetic spectrum \citep[e.g. optical photometric variations, H$\alpha$ emission, or gyrosynchrotron radio emission][]{2019MNRAS.487..304D,2020MNRAS.499.5379S,2021MNRAS.507.1979L}. Of particular interest to Polstar, the red-shifted absorption feature is a diagnostic of infalling plasma  \citep{erb21}. As can be seen in the lower panel of Fig.~\ref{fig:mhd_uvadm_compare}, which compares synthetic UV line profiles computed using an ADM (solid lines) versus an MHD (dashed lines) magnetospheric model, red-shifted absorption in the line profile is a subtle feature in the total line profile \citep[see Figures 5-6 of][for a more detailed description of the red-shifted absorption feature]{erb21}. Detection of such infalling material using e.g. Polstar will require high $S/N$ and high spectral resoluton.

\section{Linear spectropolarimetry in the UV}\label{sec:lin_specpol}

The basic idea of linear spectropolarimetry is rather straightforward: electrons in an extended circumstellar medium scatter radiation from the stellar surface,  
giving rise to a certain linear polarization level. 
If the sky-projected electron distribution is perfectly 
circular, the linear 
Stokes $Q$ and $U$ vectors cancel, and linear polarization
remains absent.
If the geometry is not circular, but   
an asymmetry is involved, this results in some level of continuum linear polarization of order 1\% \citep[e.g.][]{2002MNRAS.337..341H}.  

\begin{figure}[ht]
    \centering
\includegraphics[width=0.48\textwidth]{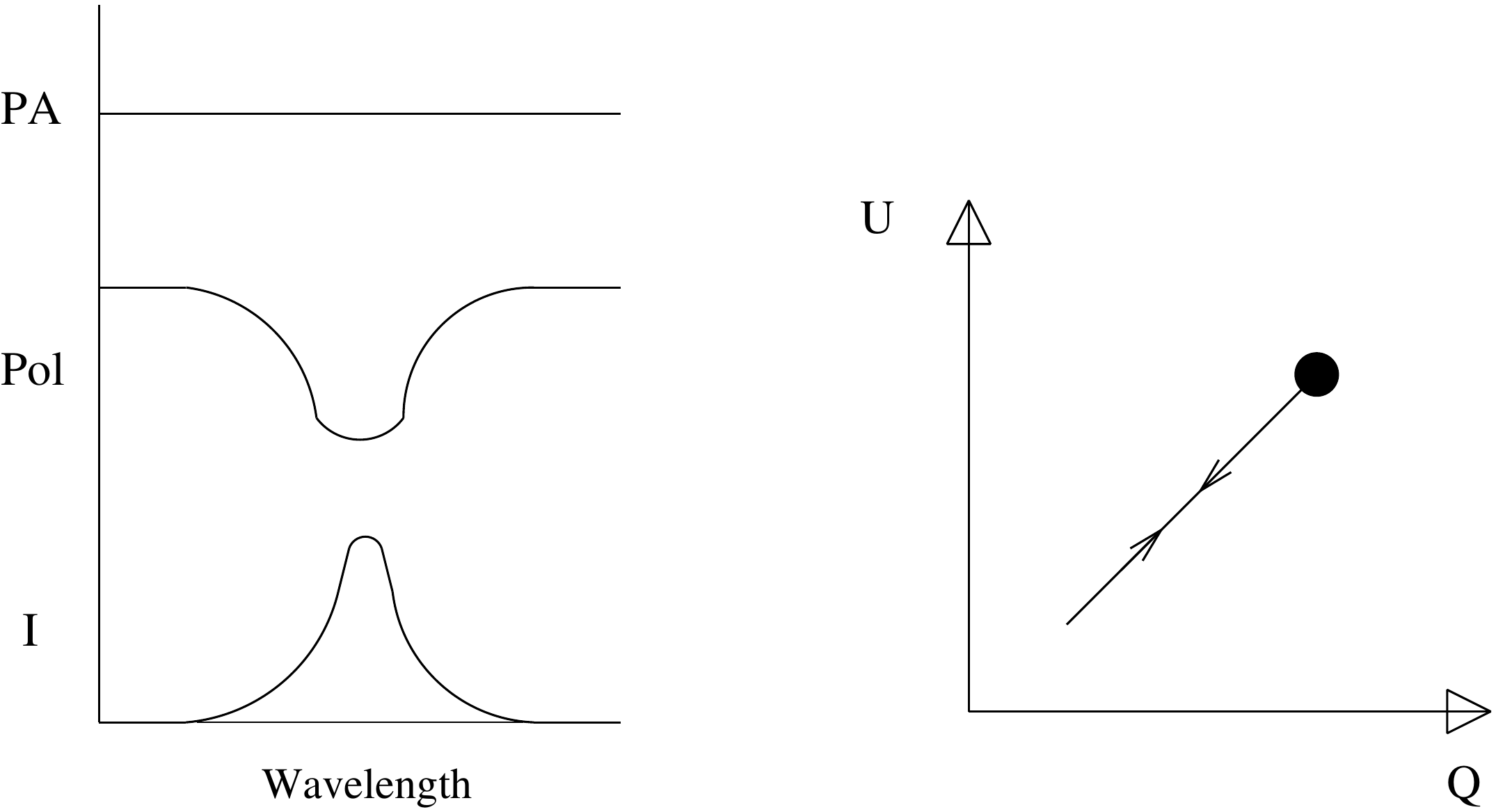}
\includegraphics[width=0.48\textwidth]{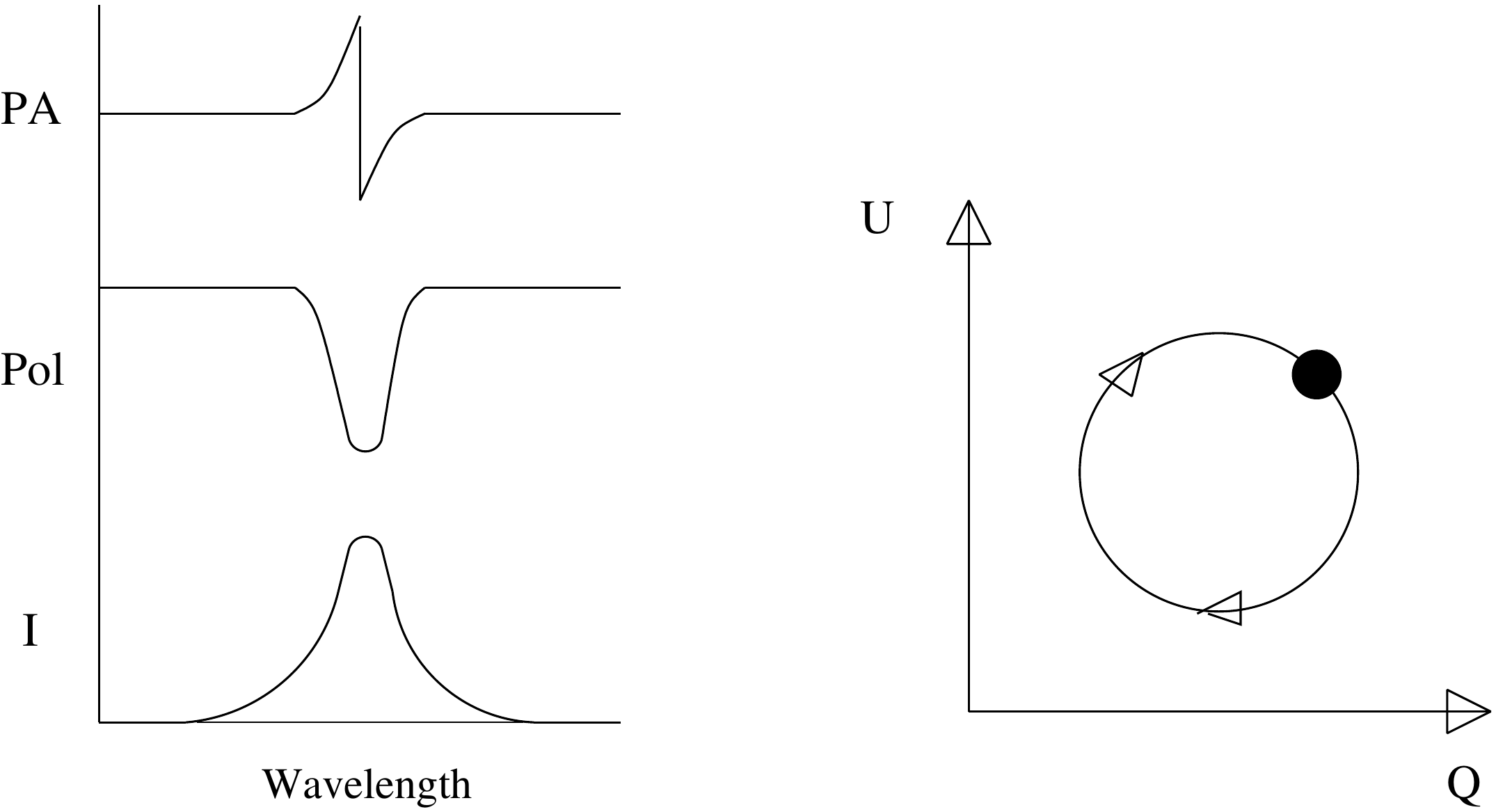}
    \caption{Cartoons representing 
line {\it depolarization} (top row) and compact line emission
scattered off a rotating disk (bottom row) as triplots and $QU$ diagrams. Stokes~$I$
profiles are plotted in the lower triplot panels, \% Pol in the middle panels, and 
the position angles (PAs) are shown in the upper triplot panels. Line depolarization is as broad as the Stokes I emission, while {\it line} 
polarization is narrow by comparison. Depolarization translates 
into $QU$ space as a linear excursion, 
while {\it line} polarization PA flips are associated with $QU$ loops.}
    \label{fig:linepol}
\end{figure}

One of the advantages of linear spectropolarimetry over continuum polarimetry is 
that it is possible to perform differential measurements between a line and the 
continuum -- independent of interstellar and /or instrumental  
polarization.
One example may occur across an emission line. This ``line effect'' relies on the expectation 
that recombination lines arise over a much larger volume than the 
continuum, and becomes {\it de}polarized (see the left hand side of 
Fig.\,\ref{fig:linepol}). 
Depolarization immediately indicates the presence or absence of 
asphericity. 

The bulk of linear spectropolarimetric studies involved these depolarization line effects, but in some situations there is 
evidence for intrinsic {\it line} polarization, such as was predicted by \cite{1993A&A...271..492W} and 
found observationally in pre-main sequence T Tauri and Herbig Ae/Be stars \citep{2002MNRAS.337..356V,2005A&A...430..213V}.
In such cases the line photons are assumed to originate from a more compact source (e.g.
as a result of magnetospheric phenomena), and these photons are scattered off a rotating 
disk, leading to a flip in the position angle (PA), resulting in a rounded loop (rather than a linear excursion) 
in the $QU$ diagram (sketched on the right hand side of Fig,\,\ref{fig:linepol}).

\begin{figure}[ht]
    \centering
\includegraphics[width=0.23\textwidth]{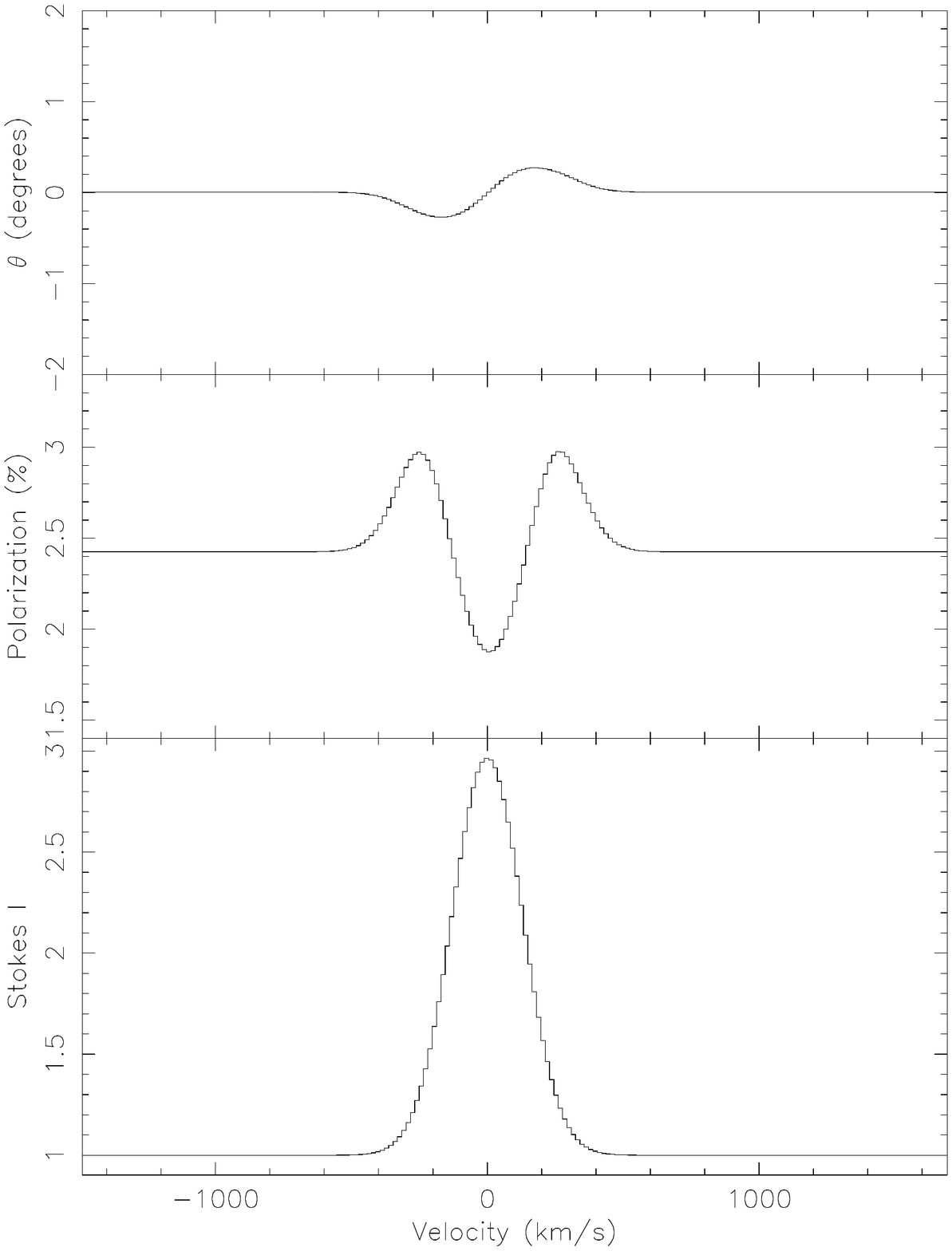}
\includegraphics[width=0.23\textwidth]{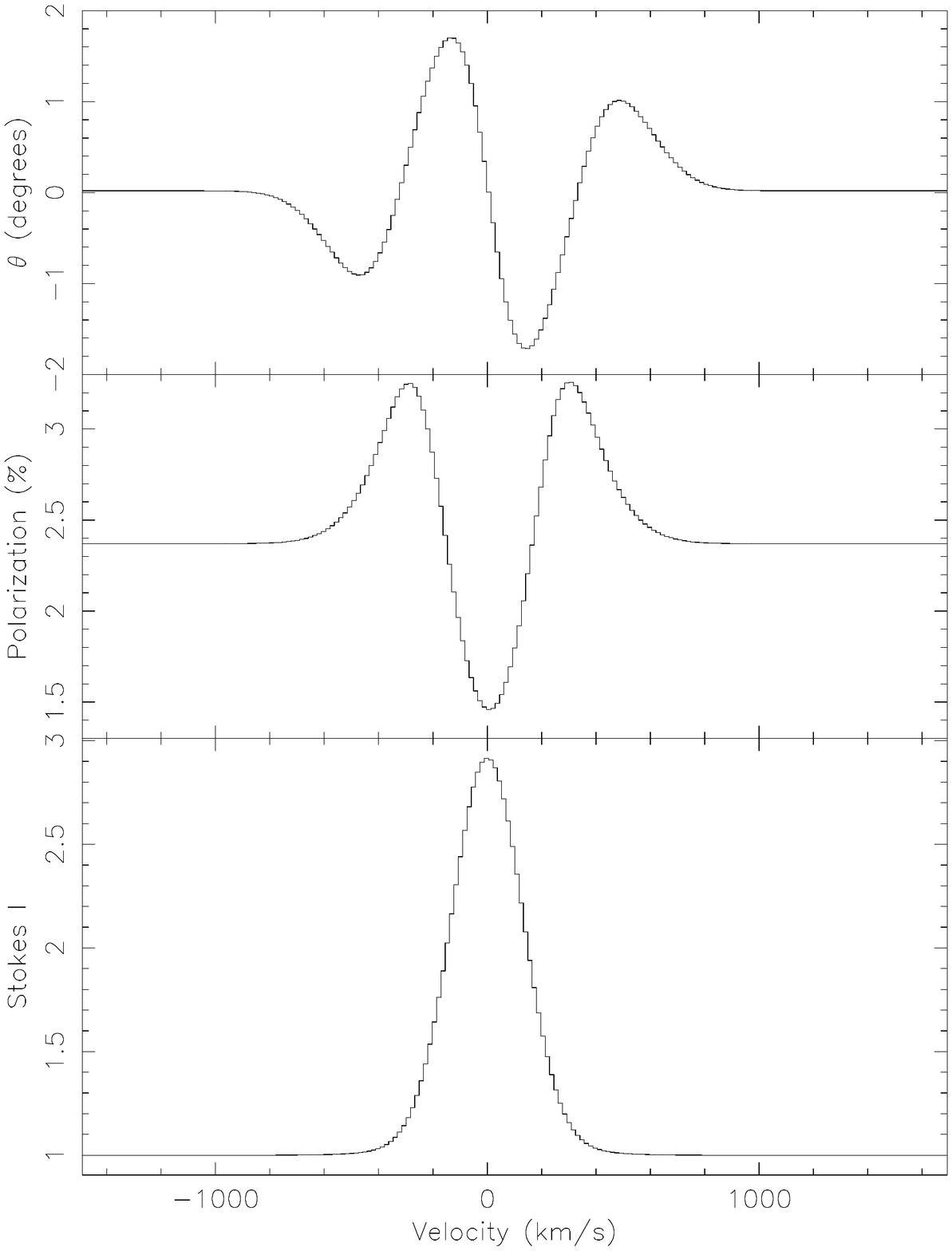}
    \caption{Monte Carlo line polarimetry for the case of a disk with an inner hole (left hand side) 
and without an inner hole -- a result of the finite size of the 
star (right hand side) \citep{2005A&A...430..213V}.}
    \label{fig:linepol-MC}
\end{figure}

Motivated by the high incidence of $QU$ loops in T Tauri and Herbig Ae stars, \cite{2005MNRAS.359.1049V}
developed Monte Carlo polarization models of scattering off rotating disks -- with and without inner holes. 
Figure \ref{fig:linepol-MC} shows a pronounced difference between scattering off a disk that hits the stellar surface 
(right-hand side), and one with a sizeable inner hole (left-hand side). 
The single PA flips on the left are similar to those predicted analytically \citep{1993A&A...271..492W}, but 
the double PA flips on the right -- associated with undisrupted disks -- have only been found numerically. They are thought to be unique to the appropriate geometric treatment of a finite-sized star that interacts with the velocity structure of the disk. 

The numerical models demonstrate the potential of instrinsic line polarimetry to determine not only disk inclination, but also the physical sizes of the inner regions associated with magnetospheres. 
Linear line polarimetry is so far the only method 
capable of doing this on such small spatial scales, within just a few stellar radii from the stellar surface.

Several studies of linearly polarized H$\alpha$ emission have been conducted for AGB, T Tauri, and Herbig Ae/Be stars by \cite{2005MNRAS.359.1049V}, \cite{2007ApJ...667L..89H,2009ApJS..180..138H}, and \cite{2016MNRAS.461.3089A,2017MNRAS.472..854A}. These have found line polarization amplitudes of order 0.1-1\%, detectable at an unbinned S/N of $\sim200-1000$. This is comparable to the S/N required for circumstellar magnetometry, therefore the same data that will enable the measurement of circumstellar magnetic fields should also enable the detection of circumstellar line polarization effects. 

\section{Continuum linear polarimetry in the UV}\label{sec:continuum_linpol}

The presence of an obliquely rotating magnetosphere is largely responsible for the periodic modulation of the observable quantities of magnetic massive stars. As the star rotates, its own magnetosphere periodically occults the source star’s light. For hot stars with winds primarily dominated by the electron scattering opacity, the bulk of their photometric and polarimetric variability can be estimated under the single-electron scattering approximation. In this case, the photometric variability is determined by the column density, while the polarimetric variability is characterised by the general shape of the magnetosphere. 

Broadband linear polarimetry of stellar magnetospheres is observationally challenging: the typical levels of polarization are quite low, of order $10^{-4}-10^{-3}$ \citep[e.g.][, and Munoz et al. (accepted)]{2013ApJ...766L...9C,munoz20}. As a consequence, accurate monitoring of magnetospheric polarization on the relevant rotational timescales (ranging from days to years) requires understanding and eliminating competing instrumental effects. 

\begin{figure*}[t]
\centering
\includegraphics[width=7.0cm]{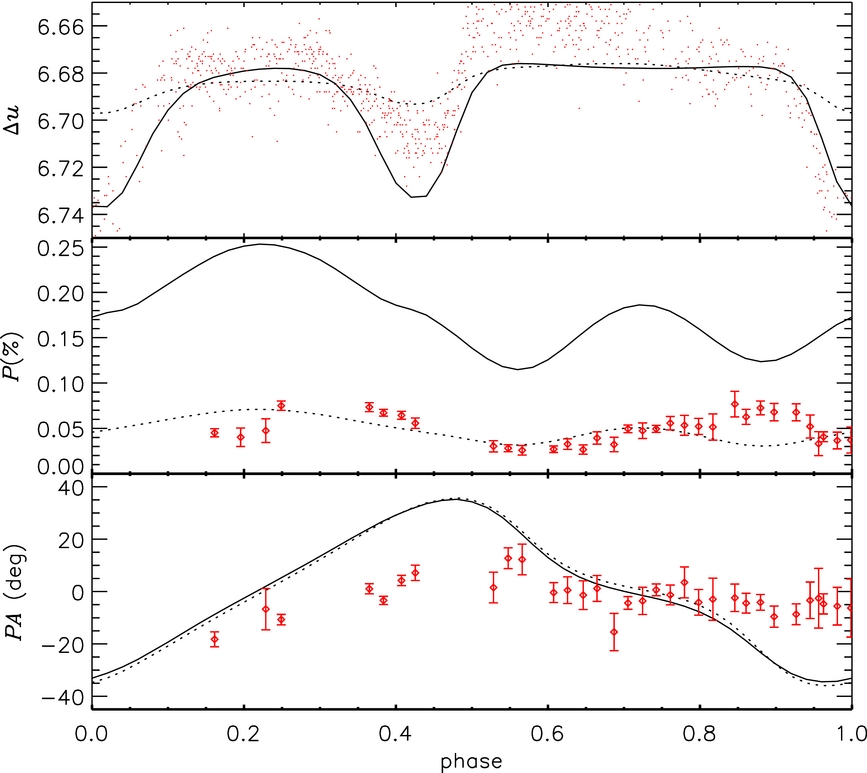}\hspace{0.5cm}
\begin{minipage}[b]{0.4\linewidth}
\centering
\includegraphics[width=7.0cm]{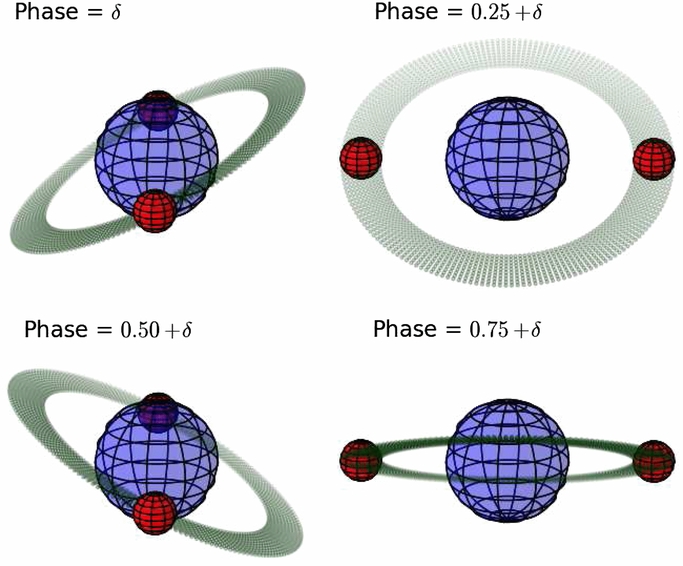}\vspace{0.4cm}
\end{minipage}
    \caption{{\em Left:\ } Modeling of the intrinsic polarization of $\sigma$ Ori E using the RRM model (observations are in red). The only free parameter is the maximum number density in the magnetosphere, which was set to $10^{12}$ cm$^{-3}$ (solid lines) to reproduce the depth of the eclipses and $2.5\times 10^{11}$ cm$^{-3}$ (dotted lines) to reproduce the amplitude of the linear polarization.  {\em Right:\ } Geometric conception of the "dumbbell + disk" model to scale. From \citet{2013ApJ...766L...9C}.}
    \label{magpol1}
\end{figure*}

Even in the optical, few studies of magnetospheric linear polarization have been carried out. The first such studies investigated the Centrifugal Magnetosphere of the archetypical He-strong Bp star $\sigma$~Ori E by \citet{1977ApJ...218..770K} and \cite{2013ApJ...766L...9C}. As shown in Fig.~\ref{magpol1}, Carciofi et al. obtained dense polarimetric coverage of the star's 1.19-day rotational cycle, measuring both the $Q$ and $U$ Stokes parameters with a typical precision of $10^{-4}$ (0.01\%, about an order of magnitude better than in the pioneering study by \citealt{1977ApJ...218..770K}). They attempted to reproduce the observed polarimetric variations by feeding the density distribution for $\sigma$~Ori E computed using the Rigidly Rotating Magnetosphere (RRM) model into a radiative transfer code. They were unable to find a model capable of simultaneously fitting both the photometry and the polarimetry, noting that a higher density model (solid line in Fig.~\ref{magpol1}) that matched the depth of the photometric eclipses predicted a polarization amplitude much larger than observed, while a lower-density model (dashed line in Fig.~\ref{magpol1}) that reproduced the amplitude of the polarization failed to reproduce the photometric amplitude. They were able to resolve these descripancies using an ad hoc ``dumbbell + disc" model with a density distribution motivated by the RRM predictions. This study serves as an excellent illustration of the power of polarization to test theoretical models of magnetospheric density and geometry.

\begin{figure*}[t]
    \centering
    \includegraphics[width=8.0cm]{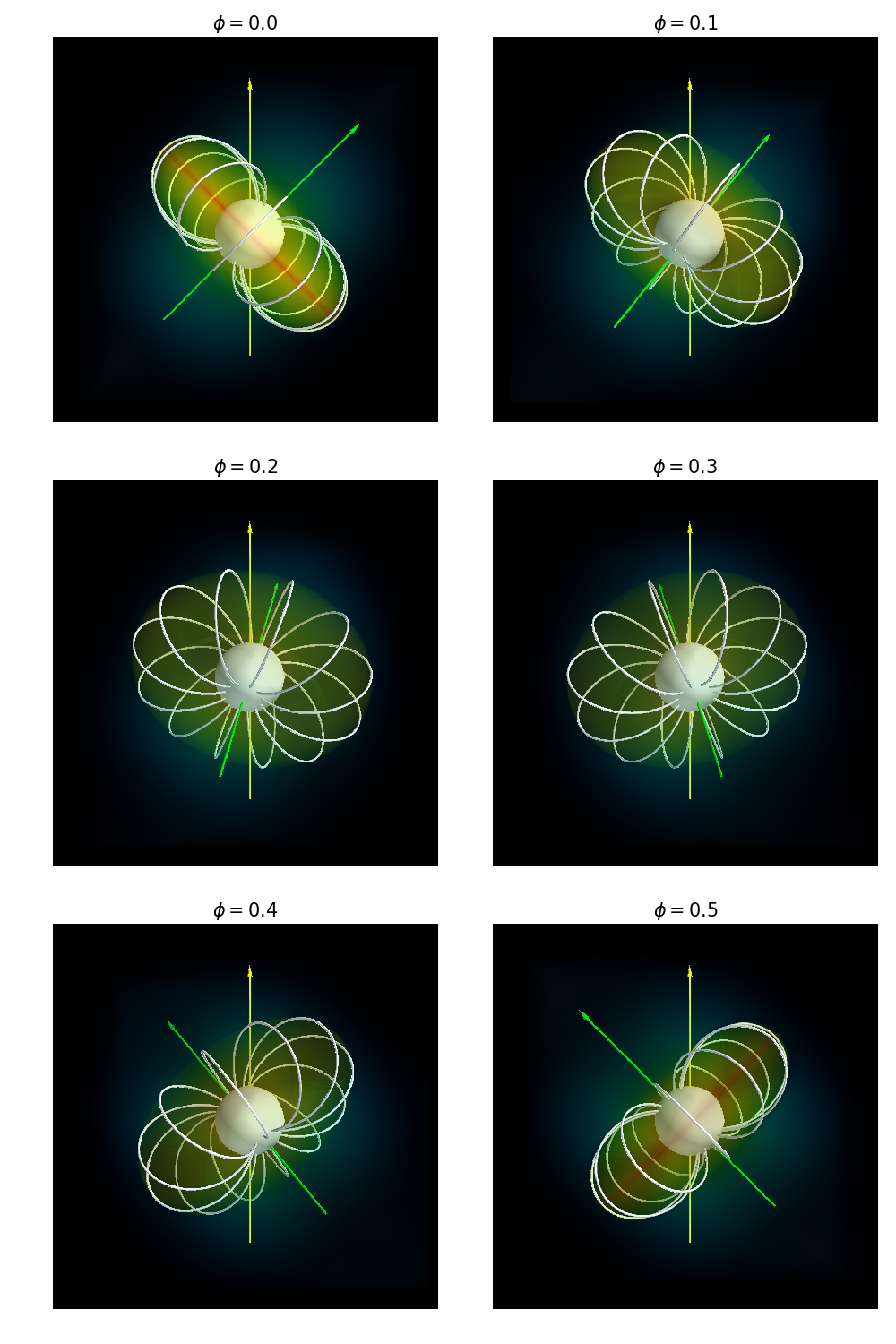}
\begin{minipage}[b]{0.4\linewidth}
\centering
\includegraphics[width=7.0cm]{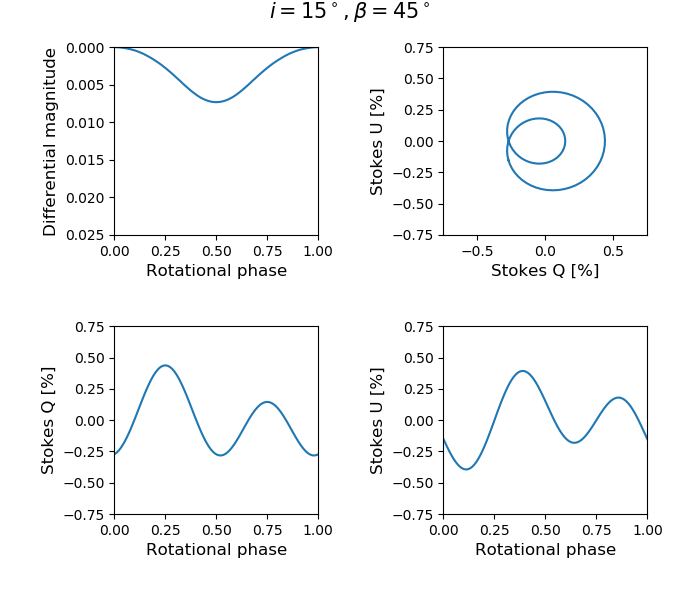}
\includegraphics[width=7.0cm]{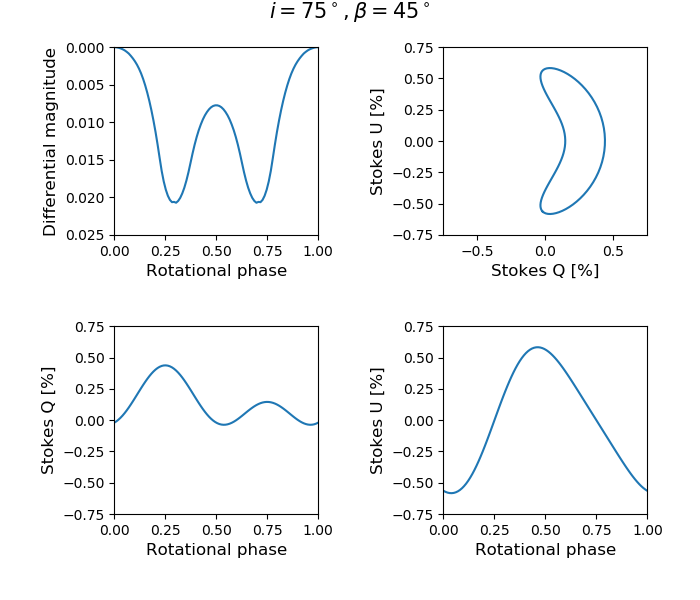}
\end{minipage}
    \caption{{\em Left:\ } Illustration of the density field of the Analytic Dynamical Magnetosphere (ADM) model at 6 rotational phases. {\em Right:\ } Photometric and polarimetric phase variations predicted for ADM models having two different geometries  ($i=15^{\circ}$, $\beta=45^{\circ}$ (top), $i=75^{\circ}$, $\beta=45^{\circ}$ (bottom)). From \citet{munoz20}.}
    \label{magpol2}
\end{figure*}

More recently, \citet{munoz20} and Munoz et al. (accepted) have developed a capability to compute magnetospheric polarization in the framework of the ADM model (Fig.~\ref{magpol2}). Under the assumption of single electron scattering, they examined the behaviour of Stokes $Q$ and $U$ with changing magnetospheric (field strength and geometry) and stellar (mass, radius, and wind) properties. They demonstrated that linear polarization is uniquely able to disentangle the angular parameters $i$ and $\beta$ describing the magnetic geometry. They applied their model to the magnetic O star HD\,191612, obtaining a self-consistent solution to the magnetic and stellar parameters capable of simultaneously reproducing the polarimetric measurements and Hipparcos photometry.

Today, high precision broadband polarization measurements are available for only a handful of magnetic stars. The potential of Polstar to collect similar data for a significant fraction of the population of known magnetic objects is extremely exciting, as it will accurate determination of their magnetic and wind parameters, and provide a broad parameteric basis for testing of the assumptions underpinning magnetospheric models. From the models and observations that have so far been published, polarization amplitudes of the order of 0.01\% to 0.1\% are expected \citep{2013ApJ...766L...9C,munoz20}.

\section{Enabled Science}\label{sec:enabled}

Massive OB stars show variability both spectroscopically and photometrically in many wavelength bands including the UV, optical and X-ray. Regular variability could be related to stellar pulsations, rotation modulation, and/or the presence of globally organized magnetic fields, while widespread stochastic variability on hourly to daily timescales is poorly understood. Stochastic variability could be caused by the presence of bright spots (possibly of magnetic origin), clumps, internal gravity waves \citep{bowman2019, bowman2020}, or subsurface convection \citep{cantiello2021}. 
A well-known example of irregular variability in UV wind-line profiles are discrete absorption components (DACs). The formation of these wind structures are ascribed to corotating interacting regions (CIRs, \citealt{mullan1984, mullan1986}), which arise above bright spots at the stellar surface \citep{cranmer1996}. The formation of bright surface spots can be caused by magnetic fields produced by dynamo action in the subsurface convection layer and brought up to the surface by buoyancy \citep{cantiello2011}. The lifetime of these fields is determined by the relatively short turnover time of the subsurface layer. The expected strength of these fields imply the local magnetic confinement parameter to be around unity \citep{udDoula2002} i.e.\ the fields are able affect the wind dynamically. 

The small-scale magnetic fields that appear at the surface as bright spots can be a driver of many surface related phenomena observed in OB stars. Short-lived bright spots give rise to low-amplitude ($\simeq 10$ mmag) photometric variability, as observed by space-based high-precision photometry (MOST, BRITE, TESS) in the light curves of many OB stars. \cite{ramiaramanantsoa2014} simulated the MOST light curve of the O7.5 III(n)((f)) star $\xi$ Per with several corotating bright spots, presumably of magnetic origin, with random starting times and lifetimes up to several rotations varying from one spot to another. Analysing simultaneous photometry from the BRITE constellation and ground-based spectroscopy for the O4I(n)fp star $\zeta$ Pup, \cite{ramiaramanantsoa2018} established an empirical link between wind structures and photospheric activity, implying a photospheric origin. Two types of variability were found: one single-periodic non-sinusoidal component with 1.78~d period, superimposed on a stochastic component. The periodic component is consistent with rotational modulation arising from evolving bright spots that were mapped at the surface. The signal was also found in the He II $\lambda$4686 wind emission line, showing signatures of corotating interaction regions that turn out to be driven by the bright photospheric spots observed by BRITE. The spots are suggested to originate from small-scale magnetic fields generated through dynamo action within a subsurface convection zone \citep[e.g.][]{cantiello2011,2020ApJ...900..113J}. The stochastic component observed in He II $\lambda$4686 line correlated with the amplitudes of stochastic light variations. Stochastic He~{\sc ii} variability was attributed to wind clumps, while stochastic photometric variability was proposed to be the photospheric drivers of the clumps.

\cite{daviduraz2017} investigated whether bright spots compatible with recent observation could be responsible for DACs observed in the UV line profiles of the O7.5 III(n)((f)) star $\xi$ Per. The authors successfully reproduced the behaviour of DACs with 4 equally spaced equatorial spots with angular radius $10^{\circ}$ and $20^{\circ}$. In this scenario spots appearing and disappearing randomly over time, with varying strengths and sizes, might explain the cyclical nature of DACs. 
Semi-analytic analysis for the spot size and amplitude needed to produce an overloaded wind that form the DAC was done by \cite{owocki2018}.

Unexpectedly, the signature of DACs has been detected in the X-ray range ($\zeta$\,Pup - \citealt{Naze2018, Nichols2021}, $\lambda$\,Cep - \citealt{Rauw2015}, $\xi$\,Per - \citealt{Massa2019}). Their presence and characteristics were not predicted by current models. UV data can however constrain in detail the spots and wind structures, leading to better modelling and a full understanding of these high-energy features.

Based on spectroscopic analysis of the He II $\lambda$4686 line for the O6I(n)fp star $\lambda$ Cep, it was proposed that cyclical variations are caused by the presence of multiple, transient, short-lived, corotating magnetic loops -- so-called {\it stellar prominences} -- likely associated with bright surface spots \citep{sudnik2016}. The prominences are represented as corotating spherical blobs of emitting gas attached to the surface of the star. Depending on its location, the blob contributes to the line profile as an emission or absorption feature. 

\begin{figure}[ht!]
 \begin{minipage}[b]{0.44\textwidth}
 \centering
  \includegraphics[width=0.9490\textwidth]{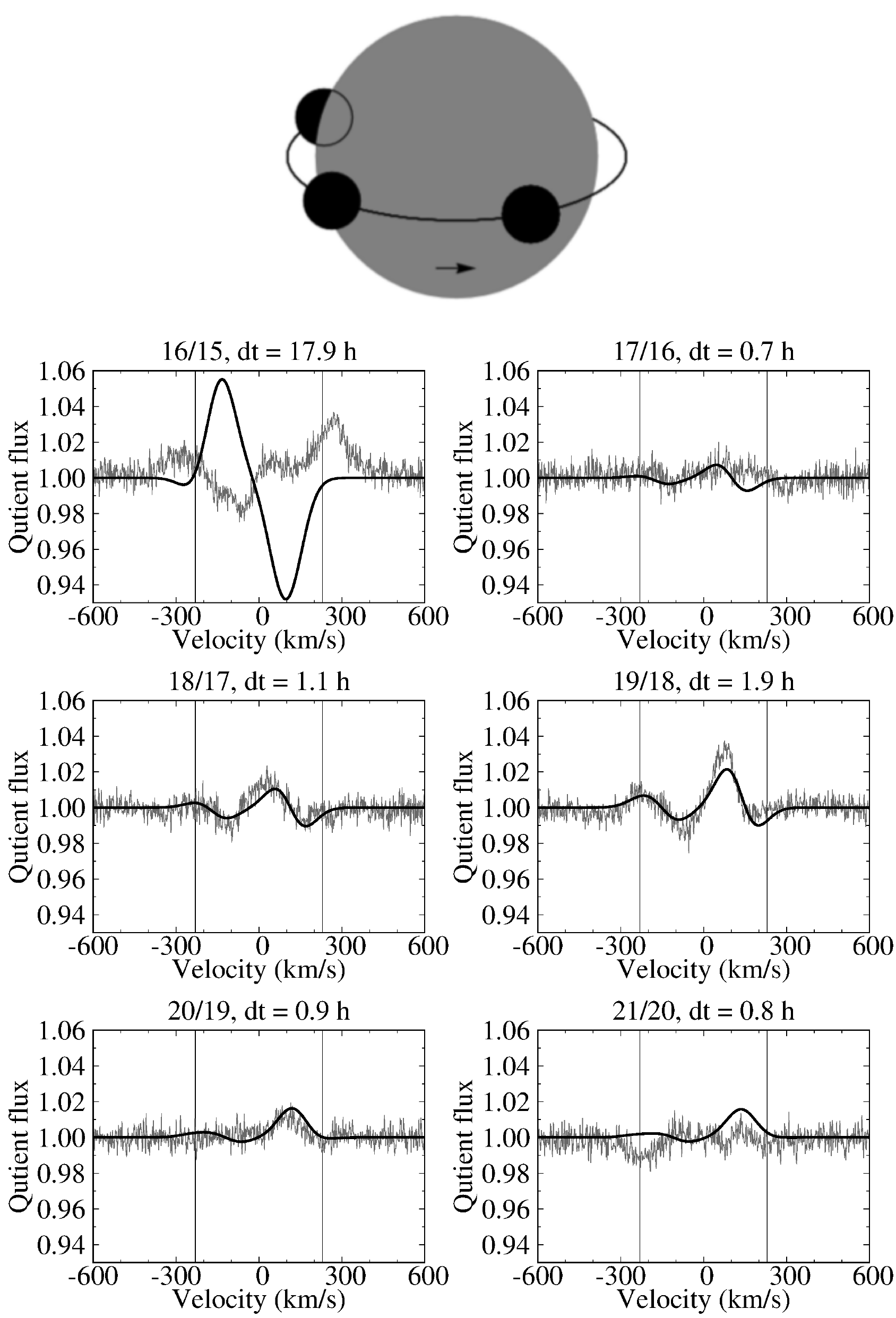}
   \end{minipage}
 \caption{Model fits (black thick lines) for subsequent quotient spectra of He II $\lambda$4686 line for the star $\lambda$ Cep (gray thin lines) of the 2013 dataset, spectra 15--21. The top label gives the sequence numbers of the two spectra of the quotient, followed by their time interval in hours. The geometry which is depicted at the beginning of the series, is used for all figures in the sequence and shown for the epoch of the first spectrum in the series. The star rotates but the blobs keep their same relative position. The first and last figure of the series contains intentionally failed fits, to signify the extend over which the fitted configuration, carried around by the rotation, survives \citep{sudnik2016}.}
   \label{stel_prom}
   \end{figure}

An example of the model fit of subsequent quotient spectra i.e., the following spectrum divided by the previous one, is shown in Fig.~\ref{stel_prom}. The proposed model applied to the \ion {He} {II} $\lambda$4686 line can be fitted with 2-5 equatorial blobs with lifetimes between $\sim$1 and 24 h. The action of the subsurface convection zone would be the most likely driving mechanism that generates short-lived magnetic bright spots as the source of prominences.

\begin{figure*}[ht!]
  %\begin{minipage}{\textwidth}
  %\centering
  \includegraphics[width=0.49\textwidth]{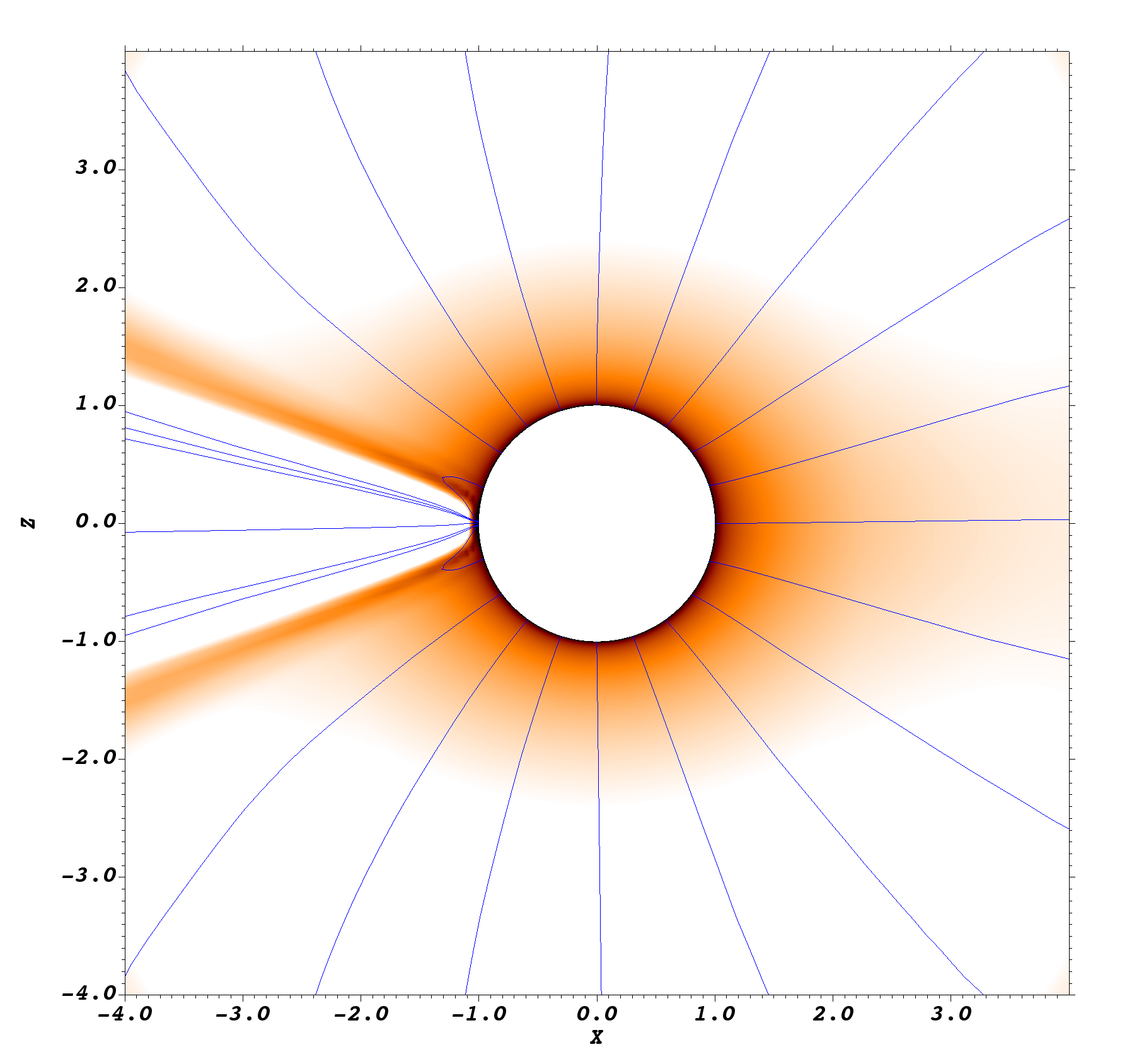}
  %\end{minipage}\hfill
  %\begin{minipage}{\textwidth}
  %\centering
  \includegraphics[width=0.49\textwidth]{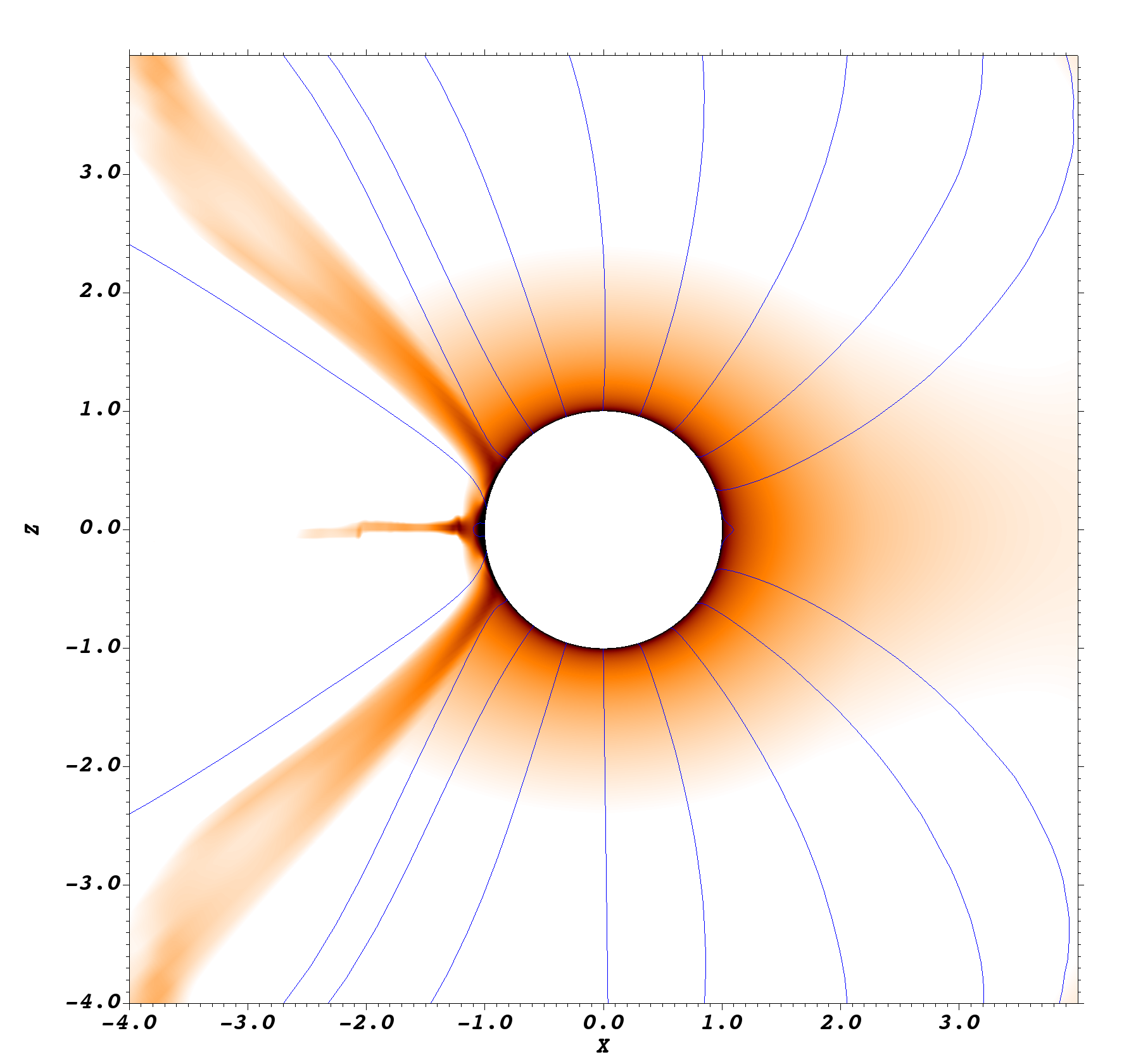}
 % \end{minipage}
  \caption{Logarithm of density of ZX-plane of 3D MHD simulations of a sample O-star wind with a small scale off-center magnetic dipole field to mimic small scale strong magnetic field that can perturb the wind at the base. The angle between rotation ($W=0.5$ half critical) and magnetic axes $\beta=0^{\circ}$ ({\it left panel}) and $\beta=90^{\circ}$ ({\it right panel}). The color code represents the logarithm of density in a range $10^{-15}$ to $10^{-11}$\,g/cm$^3$. Blue lines represent time-evolved magnetic field lines which are strong only over limited area on the stellar surface and fall of steeply with distance away from the star.}
  \label{tiny_dipole}
\end{figure*}

Local small scale magnetic fields tied to stellar prominences can be viewed as an off-center small (tiny) dipole with its center located close to the stellar surface. Such small scale fields can cause variability at the stellar surface that can propagate outwards affecting what is observed in these winds. As a proof-of-concept, we have carried out some preliminary 3D MHD simulations of hot star winds wherein we assumed that the location of the strong local field coincides with bright spots using the approach of \cite{owocki2018}.  Fig.~\ref{tiny_dipole} shows the logarithmic density of two such sample models for a typical hot star viewed along  ZX-plane. The left panel shows a case where the small scale dipole field is aligned with rotation axis, and the right panel has an inclination angle of 90$^{\circ}$ with respect to the rotational axis. We assumed rotation of 0.5 critical, and a field strength of order kG at the stellar surface that falls off steeply with distance from the photosphere. As seen in Fig.~\ref{tiny_dipole}, despite being localized, this field can substantially influence the wind dynamics globally.

While such small-scale fields cannot be detected with existing ground-based facilities, ultraviolet spectropolarimetry may provide an avenue for direct detection, for instance via the Hanle effect (if they are relatively weak), or even via the Zeeman effect (if the spots are large, the field strong, with an advantage being gained by the greater brightness of the spots in the UV). Linear spectropolarimetry will further provide important geometrical data that may help distinguish between CIR models with and without magnetic fields.

\section{Summary}\label{sec:summary}

In this  paper we have described how the unique capabilities offered by Polstar will lead to fundamental advances in our understanding of the magnetospheres of hot stars. While the focus of this  paper has been on the capabilities of the Polstar mission, this work is of relevance to any ultraviolet spectropolarimetric mission, such as Arago. 

The high-resolution ultraviolet spectra obtained by Polstar will enable much more precise spectroscopic evaluation of stellar magnetospheres, as compared to the lower-resolution, lower-$S/N$ data available for most stars via the Interstellar Ultraviolet Explorer. Importantly, over half of the non-magnetic stars proposed to be observed by Polstar do not have a single UV observation; of those that do, only a small fraction have time-series data adequate for evaluation of the projected magnetospheric geometry and column density across a rotational cycle. 

While surface magnetic field measurements obtained via ground-based visible spectropolarimetry are already available for all stars in the sample, the large number of spectral lines available for multi-line analysis in high-resolution ultraviolet spectra more than compensates for the weaker Zeeman effect at shorter wavelengths, in principle enabling higher-precision magnetic meeasurements to be obtained in the UV as compared to the visible. The full-Stokes capability of Polstar, and expected advantages in the UV over the visible in the amplitude of Stokes $QU$ signals associated with the transverse Zeeman effect, mean that many of the datasets will be optimal for magnetic mapping via full-Stokes Zeeman Doppler Imaging (ZDI). Importantly, the availability of all four Stokes parameters for magnetic inversion breaks degeneracies that can affect maps obtained only in Stokes $IV$.

Polstar will enable measurement of circumstellar magnetic fields, with the projected capabilities of the instrument capable of detecting magnetic signatures originating in the circumstellar environment in a large fraction of known magnetic stars. Strong fields should be detectable via the Zeeman effect (as evaluated using state of the art magnetospheric models). 

Both high- and low- resolution linear spectropolarimetry will provide crucial and sensitive constraints on the magnetospheric geometry, enabling degeneracies between rotational axis inclinations and magnetic axis tilt angles to be broken. Importantly, the information available via linear polarization provides geometrical data that cannot be obtained via spectroscopy or photometry alone, as already revealed by the insufficiency of current magnetospheric models to simultaneously reproduce the light curve and polarimetric variation of  $\sigma$ Ori E. 

By combining the rich spectroscopic and polarimetric datasets available with Polstar observations, detailed 3D models of the circumstellar environments of a large number of magnetic hot stars can be compared against constraints on the circumstellar magnetic field, column density, velocity structure, and geometry. This will enable measurement of the escaping and magnetically trapped wind fraction of these stars across a full range of stellar, evolutionary, magnetic, and rotational parameters, thereby providing a crucial test of the expectation that magnetic fields rapidly drain angular momentum and drastically reduce the net mass-loss rates of massive stars. This will provide empirical calibration for evolutionary models incorporating rotation and magnetic fields, which will in turn provide important information for the stellar population synthesis models used to infer the mass and energy budget for the interstellar medium, expectations for the properties of post-main sequence supergiants and supernovae, and the population statistics of stellar remnants. 

\bibliography{bib_dat}{}
\pagebreak
\newpage
\section* {Statements \& Declarations}
\subsection*{Funding}
AuD acknowledges support by NASA through Chandra Award number TM1-22001B and GO2-23003X issued by the Chandra X-ray Observatory 27 Center, which is operated by the Smithsonian Astrophysical Observatory for and on behalf of NASA under contract NAS8-03060. 

M.E.S. acknowledges financial support from the Annie Jump Cannon Fellowship, supported by the University of Delaware and endowed by the Mount Cuba Astronomical Observatory. 

A.D.-U. is supported by NASA under award number 80GSFC21M0002. 

C.E. gratefully acknowledges support for this work provided by NASA through grant number HST-AR-15794.001-A from the Space Telescope Science Institute, which is operated by AURA, Inc., under NASA contract NAS 5-26555. C.E. also gratefully acknowledges support from the National Science Foundation under Grant No. AST-2009412. 

M.C.M.C. acknowledges internal research support from Lockheed Martin Advanced Technology Center. 

This material is based upon work supported by the National Center for Atmospheric Research, which is a major facility sponsored by the National Science Foundation under Cooperative Agreement No. 1852977. 

Y.N. acknowledges support from the Fonds National de la Recherche Scientifique (Belgium), the European Space Agency (ESA) and the Belgian Federal Science Policy Office (BELSPO) in the framework of the PRODEX Programme (contracts linked to XMM-Newton and Gaia).

N.S. acknowledges support provided by NAWA through grant number PPN/SZN/2020/1/00016/U/DRAFT/00001/U/00001.

\subsection*{Competing Interests}
The authors have no relevant financial or non-financial interests to disclose.
\subsection*{Author Contributions}
All authors contributed to the study conception and design. The first draft of the manuscript was written by Asif ud-Doula and all authors commented on previous versions of the manuscript. All authors read and approved the final manuscript.
\subsection*{Data availability}
Data sharing not applicable to this article as no datasets were generated or analysed during the current study.
\newpage
\onecolumn
\centering
\appendix
\section*{Affiliations}
$^{1}${\orgdiv{Penn State Scranton, 120 Ridge View Drive, Dunmore, PA 18512, US}}

%\and
%\noindent
%\author{R. Casini}
%$^{2}${\orgdiv{High Altitude Observatory, National Center for Atmospheric Research, P.O. Box 3000, Boulder CO 80307-3000, USA}}

\noindent
%\author{M. C. M. Cheung}
$^{2}${\orgdiv{Lockheed Martin Solar and Astrophysics Laboratory, 3251 Hanover St, Palo Alto, CA 94304, USA}}

\noindent
%\author{A. David-Uraz}
$^{3}${\orgdiv{Department of Physics and Astronomy, Howard University, Washington, DC 20059, USA}}
\noindent

$^{4}${\orgdiv{Center for Research and Exploration in Space Science and Technology, and X-ray Astrophysics Laboratory, NASA/GSFC, Greenbelt, MD 20771, USA}}

%\and
%\noindent
%\author{T. del Pino Alem\'an}
%$^{6}${\orgdiv{Instituto de Astrof\'isica de Canarias, E-38205 La Laguna, Tenerife, Spain}}

%\noindent
%$^{7}${\orgdiv{Departamento de Astrof\'isica, Universidad de La Laguna, E-38206 La Laguna, Tenerife, Spain}}

\noindent
%\author{C. Erba}
%$^{5}$\affiliation{Department of Physics and Astronomy, University of Delaware, 217 Sharp Lab, Newark, Delaware, 19716, USA}
%\noindent

$^{5}${\orgdiv{Department of Physics \& Astronomy, East Tennessee State University, Johnson City, TN 37614, USA}}

\noindent
%\and
%\author{C.\ P.\ Folsom}
$^{6}${\orgdiv{Tartu Observatory, University of Tartu, Observatooriumi 1, T\~{o}ravere, 61602, Estonia}}

\noindent

%\author{K. Gayley}
$^{7}${\orgdiv{Department of Physics \& Astronomy, University of Iowa, 203 Van Allen Hall, Iowa City, IA, 52242, USA}}

\noindent

%\author{R.\ Ignace}
%$^{11}${\orgdiv{Department of Physics \& Astronomy, East Tennessee State University, Johnson City, TN 37614, USA}}

\noindent

%\author{Z. Keszthelyi}
%$^{12}${\orgdiv{Anton Pannekoek Institute for Astronomy, University of Amsterdam, Science Park 904, 1098 XH, Amsterdam, The Netherlands}}
%\noindent

%\author{O. Kochukhov}
%$^{13}${\orgdiv{Department of Physics and Astronomy, Uppsala University, Box 516, 75120 Uppsala, Sweden}}
%\noindent

%\affiliation{Department of Physics and Astronomy, University of Delaware, 217 Sharp Lab, Newark, Delaware, 19716, USA}
%\noindent

%\author{Y. Naz\'e}
$^{8}${\orgdiv{GAPHE, Universit\'e de Li\`ege, All\'ee du 6 Ao\^ut 19c (B5C), B-4000 Sart Tilman, Li\`ege, Belgium}}
%\and
\noindent

%\author{C. Neiner}
$^{9}${\orgdiv{LESIA, Paris Observatory, PSL University, CNRS, Sorbonne University, Université de Paris, 5 place\\ Jules Janssen, 92195 Meudon, France}}

%\and
\noindent
%\author{M. Oksala}
%$^{16}${\orgdiv{Department of Physics, California Lutheran University, 60 West Olsen Road 3700, Thousand Oaks, CA, 91360, USA}}
%\noindent

%\author{V. Petit}
$^{10}${\orgdiv{Department of Physics and Astronomy, University of Delaware, 217 Sharp Lab, Newark, Delaware, 19716, USA}}

\noindent

%\author{R. Prinja}
$^{11}${\orgdiv{Department of Physics and Astronomy, Gower Street, London WC1E 6BT, UK }}

\noindent

%\author{P. A. Scowen}
%$^{18}${\orgdiv{NASA Goddard Space Flight Center, 8800 Greenbelt Rd., Greenbelt, MD 20771}}

%\noindent

%\author{N. Sudnik}
$^{12}${\orgdiv{Nicolaus Copernicus Astronomical Centre of the Polish Academy of Sciences, Bartycka 18, 00-716 Warsaw, Poland}}

\noindent

%\author{J. S. Vink}
$^{13}${\orgdiv{Armagh Observatory and Planetarium, College Hill, BT61 9DG Armagh, Northern Ireland}}
%\and

%\and
%\author{G.\ A.\ Wade}
$^{14}${\orgdiv{Department of Physics and Space Science, Royal Military College of Canada, PO Box 17000, Station Forces, Kingston, ON, K7K 7B4}}

\noindent

\end{document}